%% file: casa-online-survey-chi-2023.tex
\documentclass[manuscript]{acmart}
\pdfoutput=1
\usepackage{tabularx} 
\usepackage{graphicx}
\usepackage{mdframed}
\usepackage[inline]{enumitem} 
\usepackage{multirow} 
\usepackage{siunitx} 
\usepackage{tablefootnote} 
\usepackage{csquotes}
\usepackage[gen]{eurosym}

\AtBeginDocument{%
  \providecommand\BibTeX{{%
    \normalfont B\kern-0.5em{\scshape i\kern-0.25em b}\kern-0.8em\TeX}}}

\renewcommand\footnotetextcopyrightpermission[1]{} 
\newcommand{\changes}[1]{{#1}}
\newcommand{\changesAdded}[1]{}
\newcommand{\removed}[1]{}

\newcommand{\ie}{i.\,e.}
\newcommand{\eg}{e.\,g.}

\newcommand{\etal}{et~al.\@\,}

\begin{document}

\input{sections/00_preamble.tex}

\input{sections/01_abstract}

\input{sections/02_intro}

\input{sections/04_relatedwork}

\input{sections/05_methodology}

\input{sections/06_results}

\input{sections/07_discussion}
\input{sections/08_conclusion}

\bibliographystyle{ACM-Reference-Format}
\bibliography{casa-online-survey-chi-2023}

\clearpage
\appendix
\input{appendix/CountryStudies.tex}
\clearpage
\input{appendix/Model_Variables}
\input{appendix/tab-reliability-predictors}
\clearpage
\input{appendix/questionnaire}
\end{document}

%% file: sections/00_preamble.tex
\ccsdesc[300]{Security and privacy~Social aspects of security and privacy}

\title[A World Full of Privacy and Security (Mis)conceptions?]{A World Full of Privacy and Security (Mis)conceptions?\\Findings of a Representative Survey in 12 Countries}

\author{Franziska Herbert}
\affiliation{
  \institution{Ruhr University Bochum}
  \country{Germany}
}
\email{franziska.herbert@rub.de}
\author{Steffen Becker}
\email{steffen.becker@rub.de}
\orcid{0000-0001-7526-5597}
\affiliation{%
  \institution{Ruhr University Bochum}
  \country{Germany}
}
\affiliation{%
  \institution{Max Planck Institute for Security and Privacy}
  \country{Germany}
}
\author{Leonie Schaewitz}
\author{Jonas Hielscher}
\author{Marvin Kowalewski}
\author{M. Angela Sasse}
\affiliation{
  \institution{Ruhr University Bochum}
  \country{Germany}
}
\email{firstname.lastname@rub.de}

\author{Yasemin Acar}
\email{acar@gwu.edu}
\affiliation{
    \institution{The George Washington University}
    \country{USA}
}

\author{Markus Dürmuth}
\email{markus.duermuth@itsec.uni-hannover.de}
\affiliation{
    \institution{Leibniz University Hannover}
    \country{Germany}
}

\renewcommand{\shortauthors}{Herbert, Becker, et al.}

%% file: sections/01_abstract.tex
\begin{abstract}
Misconceptions about digital security and privacy topics in the general public frequently lead to insecure behavior. 
However, little is known about the prevalence and extent of such misconceptions in a global context.
In this work, we present the results of the first large-scale survey of a global population on misconceptions: We conducted an online survey with \changes{$n=12,351$}~participants in $12$~countries on \changes{four} continents. 
By investigating influencing factors of misconceptions around eight common security and privacy topics (including E2EE, Wi-Fi, VPN, and malware), we find the country of residence to be the strongest estimate for holding misconceptions. 
We also identify differences between non-Western and Western countries, demonstrating the need for region-specific research on user security knowledge, perceptions, and behavior.
While we did not observe many outright misconceptions, we did identify a lack of understanding and uncertainty about several fundamental privacy and security topics. 
\end{abstract}

\keywords{Security Misconceptions, Online Survey, Co-variance Analysis, Human-Centered Security}
\maketitle

%% file: sections/02_intro.tex
\section{Introduction}
Despite the huge amount of advice for users on staying safe and private online, users' misconceptions exist across many aspect of digital security and privacy~\cite{abu-salma-17-communication, wu-18-private-browsing, kang-15-data-everywhere, anell-20-threats, mayer-17-pw-misconceptions}.
Users’ misconceptions likely lead to less secure behavior~\cite{keil-22-german-id-fido, krombholz-19-mental-models-https}, increasing users' risk to be harmed. Understanding these misconceptions is thus considered an essential factor for providing users with accessible advice tailored to debunk misconceptions and is of critical importance to better educate users about secure online behavior~\cite{story-21-web-privacy, schaewitz-21-e2ee}. 
A multitude of usable security and privacy research focuses on advice, and advice sources~\cite{reeder-17-152-simple-steps, redmiles-20-advice, redmiles-16-advice-digital-security} as well as on whether and how users are aware of and understand certain security aspects, such as end-to-end encryption or HTTPS~\cite{krombholz-19-mental-models-https, abu-salma-17-communication}. 
Device usage and usage habits differ worldwide~\cite{tomczyk-19-digital-divide}, thus misconceptions likely also differ around the world. 
However, \changes{most of the literature on (mis)conceptions focuses on participants from Western countries. 
See \autoref{appendix:StudiesByCountries} for an overview of related work organized by misconception topic and country studied.}  
\changes{Although previous work has pointed to differences, particularly  between Western and non-Western countries (see \autoref{sec:relatedWork}),} a comprehensive overview of users' misconceptions across different topics and countries is still missing.
Our study aims to fill this research gap by investigating (mis)conceptions around the world \changes{and} shedding light on the prevalence and factors \changes{that influence} security and privacy (mis)conceptions. 

In this paper, we therefore answer the following research questions: 
\begin{description}
    \item[RQ1:] How widespread are misconceptions about digital security threats in various security and privacy\changes{-related} topics \textit{\changes{around} the world}? 
    \item[RQ2:] What \textit{factors} influence users' misconceptions about various digital security and privacy topics? 
\end{description}

To address our research questions, we conducted a quota-representative online survey with $n = 12,351$~participants in $12$~countries from \changes{four} continents \changes{representing} 42\% of the world's population~(see~\autoref{fig:map}). 
We cover a variety of cultures and countries differing in their economic status and Internet access. 
Our country selection was limited to countries for which online panel providers could offer an approximate representative sample. 
Therefore, the countries researched in this study are China, Germany, Great Britain, India, Israel, Italy, Mexico, Poland, Saudi Arabia, Sweden, the USA, and South Africa. 
By including countries around the world and \changes{various} security and privacy-related topics, we shed light on which issues and factors are relevant in which country. 
We addressed (mis)conceptions for areas like secure communication, secure browsing, device security, and authentication.

Our key findings are: 
\begin{itemize}
    \item Many users worldwide show neither high rates of agreement nor high rates of disagreement with a variety of digital security misconceptions. This indicates \textbf{general uncertainty} about these topics.
    \item Some \textbf{misconceptions are prevalent around the world}, \eg, ``It is important for the security of my user accounts to regularly change the password.''
    \item Certain \textbf{security aspects are globally recognized}, \eg, the risk of shoulder-surfing.
    \item One of the \textbf{most important factors influencing user misconceptions is the country of residence}, with greater differences between Western and non-Western countries. 
\end{itemize}

Collectively, our findings provide a first overview of (mis)conceptions \changes{about} security and privacy-related topics around the world. 
We discuss \changes{what} factors influence \changes{the various} misconception topics and which misconceptions are most prevalent \changes{in} different countries.

%% file: sections/04_relatedwork.tex
\section{Related Work}
\label{sec:relatedWork}
A number of previous studies investigated users' understanding, misconceptions, knowledge, and behavior \changes{on} digital security and privacy topics, including across different countries. 
\changes{The prior work presented in this section forms the foundation of our survey. Additionally, \autoref{sec:method} includes prior work on which we based certain parts of the questionnaire.}

\subsection{End User Understanding and Misconceptions}
Prior work on user understanding and misconceptions is diverse, with studies researching understanding of the Internet as a whole, and other studies focusing on mental models of specific aspects like HTTPS. 

In a qualitative interview study, Kang \etal~\cite{kang-15-data-everywhere} researched users' knowledge and mental models of the Internet and how they affect users' privacy and security decisions. 
Even though they found differences in mental models \changes{for} the Internet of lay\changes{people} and users with computer science (\ie, technical) background, they could not find a direct relationship between technical background and security measures taken. 
However, participants with technical background were more likely to secure their connection and scored better on technical knowledge questions regarding privacy. 
The authors also found a correlation between awareness of threats and the number of privacy protection measures, suggesting that awareness is a better predictor for protective measures than technical background. 
Additionally, users stated to refrain from using protective actions due to beliefs such as \textit{no one is interested in my data}.  
In an interview study with $59$~users, Kocabas \etal~\cite{kocabas-21-unauthorized} found that misconceptions about the protection of online accounts are widespread, especially the belief that one \textit{cannot do anything} to protect the\changes{ir} accounts.

In 2009, Klasnja \etal~\cite{klasnja-09-wifi} investigated privacy-threatening misconceptions of Wi-Fi users. Their results primarily show that users underestimated the risks \changes{of} wireless network connections \changes{at that time}, when HTTPS was not \changes{yet widespread} and thus \changes{spying on} private communication was comparatively easy.

In a qualitative interview study, Krombholz \etal~\cite{krombholz-19-mental-models-https} found that users confuse encryption and authentication, and underestimate the security benefits of HTTPS. The participants mistook a second authentication factor (\eg, for online banking) as an additional layer of encryption. Users did also not know about security indicators and reported they had never noticed them. 
In an interview study investigating user perception of end-to-end encrypted communication tools, Abu-Salma \etal~\cite{abu-salma-17-communication} also found participants to confuse encryption with authentication. Participants in this study additionally believed that encryption could be broken, especially by the service provider. Participants also perceived both text messages and emails to be more secure than instant messages. 

Story \etal~\cite{story-21-web-privacy} researched the adaption and user perception of privacy tools, such as VPN, private browsing, and Tor in the US. They found that most participants were slightly concerned about their privacy but also stated to know how to use privacy tools. They \changes{discovered} the misconception that these tools protect users from security threats. For example, participants believed that private browsing would prevent hacking, as it would make the device hard to find by hackers. 
These misconceptions can be harmful, as participants \changes{may} feel more protected than they are. However, Story~\etal also found that users had a certain understanding of privacy tools. Participants correctly stated VPN would ``mask one's IP address.''
A study by Wu \etal~\cite{wu-18-private-browsing} \changes{in} which users \changes{were exposed to} browsing scenarios found \changes{that they had} misconceptions. Participants believed \changes{that} private browsing mode would \textit{prevent \changes{disclosure of} geolocation, advertisements, viruses, and tracking by both the websites visited and the network provider}.

\subsection{Cross-Cultural Studies on Privacy and Security}
\changes{Previous} research has found that perceptions and behaviors \changes{regarding} security and privacy \changes{vary} across countries. 

Recent research on disclosure \changes{of} study context in CHI \changes{article} titles~\cite{kou-18-titling} revealed that most authors and participants come from Western countries, \changes{particularly} the United States and Europe. 
The authors found \changes{that the titles of papers} explicitly \changes{mentioning} the country were mostly from non-Western countries. 
They concluded that results from the US and Europe are \changes{considered} as the ``norm'' in our research community, \changes{while} studies from non-Western countries are viewed as ``\changes{exotic}''. 
The authors suggest to reconsider these norms. 

A study by Wang~\etal~\cite{wang-11-concerns} on privacy attitudes and practices of social network sites~(\changes{SNS}) with users from India, China, and the US found differences in privacy concerns between participants. 
They found \changes{that} Indian \changes{are the least concerned about their} privacy. 
Even though US participants were most concerned \changes{about protecting their} privacy, they \changes{were least likely} to limit \changes{the} visibility of their information on \changes{SNS}. 
Chinese participants showed the \changes{greatest} desire to restrict visibility. 

Sawaya~\etal~\cite{sawaya-17-behavior} \changes{examined} security behavior and its \changes{predictive} factors in an online questionnaire study \changes{in} seven countries:
China, France, Japan, Russia, South Korea, USA and United Arab Emirates. The study found country, income, technical familiarity (a job or degree in technical areas), self confidence, and technical knowledge to be significant predictors for security behavior. Participants' confidence in computer security-knowledge had a larger effect on users security behavior than their actual knowledge. The authors report less security behavior by participants from Asia (especially Japan) compared to participants from the other countries. 

Harbach~\etal~\cite{harbach-16-keep-on} conducted an online study on smartphone locking behavior across eight countries (Australia, Canada, Germany, Italy, Japan, Netherlands,
United Kingdom, United States). The study found that the participants from non-US countries (except for Italy) were more likely to use a secure lock screen. Additional to country, demographic factors like age, and gender were found to be significant predictors for using a lock screen. Older users were found less likely to use a secure lock screen. The authors also found differences in considering data on users smartphones as sensitive between countries. Japanese participants perceived data on their smartphone as more sensitive than participants from other countries. Harbach~\etal found the largest deviations from general result patterns in their study for a non-Western country. 

In a questionnaire study with computer science students (master's level) \changes{of} different nationalities (mostly \changes{Finns} and Chinese), Chaudhary~\etal~\cite{chaudhary-15-online-sec} researched security knowledge and attitudes. Even though all participants were IT students at the master level, they were found prone to security threats and to hold dangerous misconceptions. For example they found students to overlook important security and privacy properties like ``correctness of the URL''. Authors furthermore found \changes{that} students rely on less secure measures, \changes{such as} the \textit{look and feel} of a website, when assessing \changes{the legitimacy of a} website or email.

%% file: sections/05_methodology.tex
\section{Methodology}
\label{sec:method}

\begin{figure}[!ht]
	\centering
	\includegraphics[width=\textwidth]{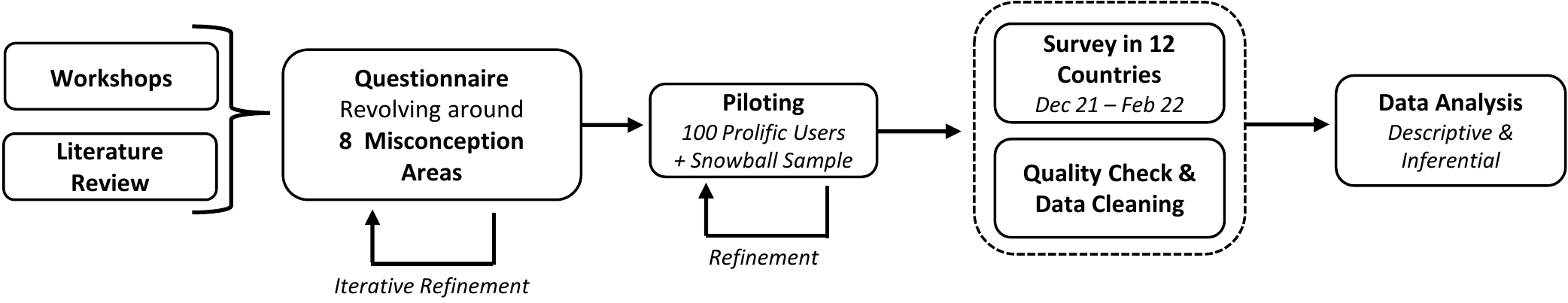}
	\Description[Method]{A description of our method, visualized as a block-flow-diagram: A combination of workshops and a literature review, is followed by the creation of a questionnaire around 8 misconception areas (that were iteratively refined). Then a piloting with 100 prolific users took place before the survey was carried out in 12 countries from December 2021 till February 2022. Data cleaning and quality was done while the survey was carried out. In a final step the data was analyzed.}
    \caption{\changes{Overview of the methodological approach and timeline of our study on privacy and security misconceptions.}}

	\label{fig:studyDesign}
\end{figure}

\begin{figure}[!ht]
	\centering
	\includegraphics[width=\textwidth]{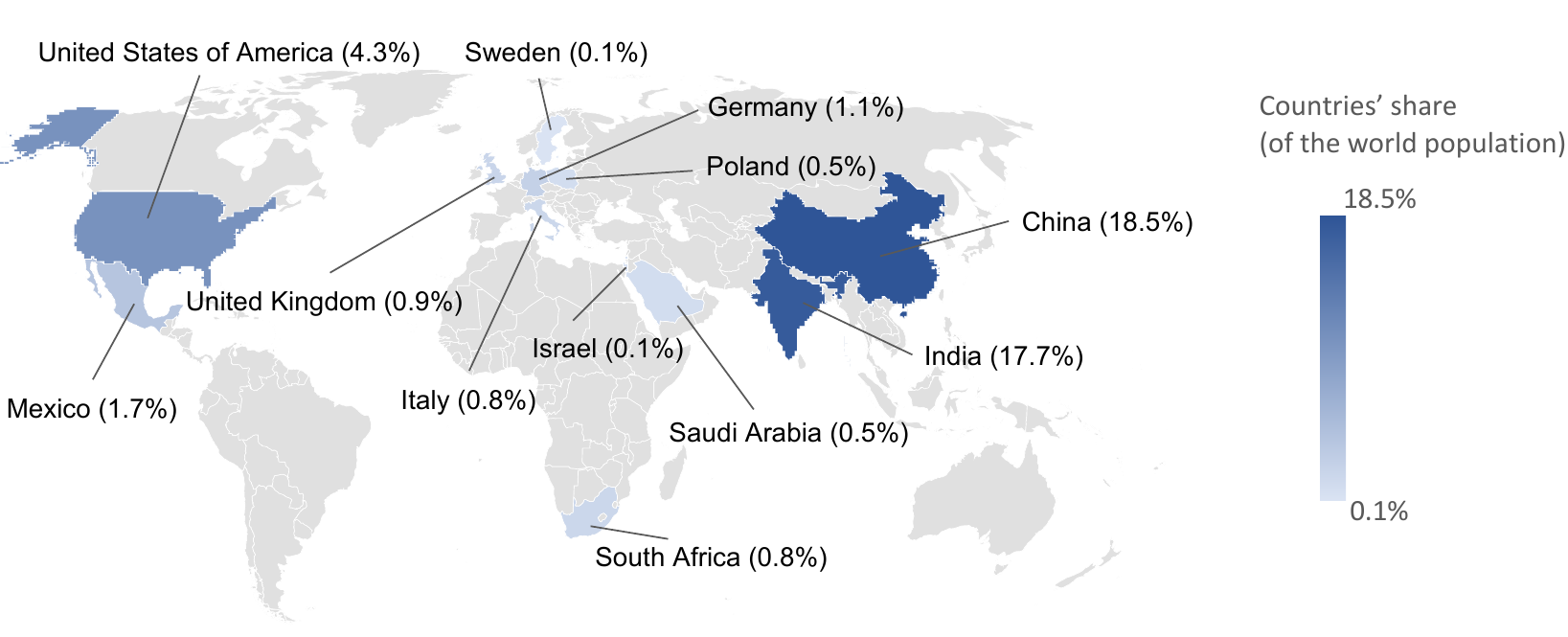}
	\Description[WorldMap]{In a world map the 12 countries we performed our survey in are highlighted and the share of the world population of those countries are displayed.}
	\caption{Online Survey in 12 countries with a representative sample of \changes{about} 1,000 participants per country (total number of $n = 12,351$ participants). The surveyed countries account for $42\%$ of the world's population. The colored legend shows each country's share of the world population, \eg, the USA has a $4.3\%$ share of the global population, placing it as the third most populous country surveyed in our online questionnaire.}
	\label{fig:map}
\end{figure}

To gain insights into users' security-relevant misconceptions, concerns and attitudes, we conducted an online survey in 12 countries on \changes{four} continents, accounting for 42\% of the world's population. 
Our sample is geographically and culturally diverse, including participants from \changes{countries such as} Great Britain, Germany, South Africa, Saudi Arabia, \changes{or} India \changes{(see \autoref{fig:map} for a complete overview on the surveyed countries)}. 
Our goal was to sample \changes{about} $1,000$ participants per country, leading to a total number of $n = 12,351$ participants.
At the core of our study, we presented participants with statements reflecting common misconceptions about digital security from eight different security-relevant topics such as authentication, device security, or encrypted communication.
\autoref{fig:studyDesign} \changes{provides an overview of our methodological approach, which is described in detail in the following sections.}

\subsection{Topic Selection and Item Generation}
\label{Misconceptions}
In the first step, we identified security threats \changes{to} users in short workshops with seven interdisciplinary security and privacy researchers. \changes{Each} researcher listed all digital security threats and advice for users they knew or researched, and we combined and summarized those lists. \changes{A subsequent} discussion \changes{of} the topics, specific advice, and threats resulted in a list of digital security and privacy aspects that \changes{carry} the possibility \changes{of} (mis)conceptions.  

We complemented our workshop results with related research on security threats for users \cite{anell-20-threats}, on cybercrime reports of different countries~\cite{zindler-20-digitalbarometer, enisa-20-threats, serianu-16-threats}, and on advice from experts for users to implement~\cite{reeder-17-152-simple-steps}. 
Based on these threats and this advice, we looked for corresponding literature mentioning user misconceptions on these topics. We focused on studies asking users about their threat models, mental models of digital security and privacy, and their beliefs about security and privacy. 
We manually clustered the topics identified in the workshops and the literature review into eight broad areas of misconceptions: 
End-to-end encrypted communication, 
four aspects related to surfing the Internet (HTTPS, Wi-Fi, VPN, private browsing mode), password and login processes, device security, as well as malware and deception. 
For each of these eight topics, we based some of our misconception statements on prior research on the respective topic and also included self-generated statements. 
Prior research, such as the studies by Kang \etal \cite{kang-15-data-everywhere}, Story \etal \cite{story-21-web-privacy}, and Anell \etal \cite{anell-20-threats} presented a variety of misconceptions and were therefore used as a foundation for different misconception statements across topics. 
For each topic, we also included correct statements which represented correct functioning of the \changes{respective} tool or measure, like ``When I use a VPN, my internet provider can no longer see what websites I visit.'' Based on previous research findings and applying advice on questionnaire design and wording~\cite{moosbrugger-12-questionnaires, clark-95-validity, redmiles-17-best-practices}, we carefully formulated most of the questions and statements in the questionnaire ourselves. 


\subsubsection{End-to-end Encrypted Communication} 
One efficient measure to protect communication is to implement end-to-end encryption for emails and for messenger services. 
For this topic, we based our misconception statements mainly on the findings of Abu-Salma \etal \cite{abu-salma-17-communication}, targeting especially how and from whom an end-to-end encrypted message is protected. An example statement is: ``If messages are end-to-end encrypted, they can also be read by third parties during transmission.'' We also included correct statements like: ``If my chat messages are protected by end-to-end encryption, then my messages can only be read on my device and by the recipient; nobody else can access and read them in transit.'' This topic consisted of nine statements. 

\subsubsection{HTTPS} Misconceptions about \textit{HTTPS} were generated on the basis of studies by Krombholz~\etal~\cite{krombholz-19-mental-models-https} and Story~\etal~\cite{story-21-web-privacy}. We generated five HTTPS-related statements for our questionnaire. Again, we included misconceptions like ``If I visit websites that use HTTPS then other people that use my computer cannot see where I have been on the Internet'', as well as true statements about HTTPS like ``If HTTPS is used on a website, my Internet provider does not know what I am clicking on the website.''

\subsubsection{Wi-Fi} 
The misconception statements relating to surfing the Internet with a special focus on \textit{Wi-Fi} were inspired by the findings by Klasnja \etal~\cite{klasnja-09-wifi}. One main threat mentioned by users was the hacking of their computers through Wi-Fi. User also thought that hackers were able to see what the user sees. Although users in this study \changes{felt that} these actions \changes{were} not very common, 
we \changes{rely on this study for} some of the five misconception statements \changes{about} Wi-Fi. One example is ``When I use a public Wi-Fi, other devices that are also using this Wi-Fi can generally see what data (\eg, passwords, credit card information) I enter on websites.''

\subsubsection{VPN} 
Our questionnaire also included misconception statements concerning surfing the Internet with special focus on \textit{VPN and the Tor browser}, as those are effective privacy tools. Story \etal~\cite{story-21-web-privacy} found that users think privacy tools, like VPN, also protect them from security risks. 
More than half of the users in their study thought VPN's protect them from hackers gaining access to their devices. Therefore, we included eight misconception statements like ``A VPN protects me from unauthorized persons getting access to my device.'' Again, we also included true statements such as ``Surfing via the Tor network prevents my Internet provider from seeing what websites I visit.''

\subsubsection{Passwords and Login Processes} 
The \changes{digital security} topics users are most \changes{likely to face are} \textit{password and login processes}, as many devices and services require authentication methods. Therefore, users face a lot of advice and myths relating to secure authentication, not only in a work environment, but also in daily life, \ie, password policies when setting up accounts. We generated 17 statements for this topic. Our statements were inspired by the systematic literature review by Mayer and Volkamer~\cite{mayer-17-pw-misconceptions}. They identified $23$ misconceptions about password security, \eg, that a word from another language or someone else's date of birth \changes{would be} a secure password. We based some statements on these findings, \eg, ``A date of birth is a secure password as long as it isn't my own date of birth.''. We also included statements about biometric authentication, as these methods are \changes{nowadays} widely used. Statements about password managers were included, as they help users to store and generate secure passwords. As for all topics, we also included correct statements \eg, ``Password managers generate secure passwords that cannot be guessed, even with technical assistance.''

\subsubsection{Device Security} 
Under the topic \changes{of} \textit{device security}, we subsume measures \changes{that} users take to secure their various devices, \changes{\eg}, using anti-virus software and \changes{their} updating behavior. We based our nine statements for this topic on a variety of papers~\cite{anell-20-threats, mayer-17-pw-misconceptions, kocabas-21-unauthorized, reeder-17-152-simple-steps} that reported on measures to secure devices. They, for example, found users to be afraid of physical theft of their devices.  
One example of our misconception statements is ``Even if my laptop is stolen, my data is secure because my user account is protected by a password.'' We also included true statements like ``To protect the data on my laptop even if it is stolen, a hard drive encryption must be used.''

\subsubsection{Malware and Deception on the Internet} 
Misconceptions related to malware and deception on the Internet included statements about how malware can be spread and the damage it may cause, but also about phishing and malicious websites. Our 16 \textit{malware and deception-}related statements were derived from literature about malware myths~\cite{willems-19-myths}, and from literature on the trustworthiness of websites and user interaction with phishing~\cite{kirlappos-12-phishing, dong-08-phishing}. One malware myth we integrated is ``If I don't discover anything suspect on my computer, then it is not infected with malware.'' Regarding phishing and fake websites, one example statement is ``As long as a website looks official, I can enter my login data without concern.'' We also included true statements like ``Links in emails can lead to fake websites to gather my login data.'' 

\subsubsection{Private Browsing} 
We generated six misconception statements related to surfing the Internet with a special focus on \textit{private browsing}.  Some of these statements were based on the misconceptions described by Wu \etal \cite{wu-18-private-browsing}. The authors state that users think \changes{that} private browsing mode could protect them from, \eg, viruses, advertisements, and tracking. Therefore, one of the statements in our survey is ``The private browser mode prevents malware from reaching my device.'' Again, we included also true statements like ``The private browser mode protects me from other people using my device from being able to track my activities.'' 

\subsection{Questionnaire Design} \label{Questionnairedesign}
\changes{In addition} to misconceptions, our questionnaire asked a \changes{number} of questions \changes{about various} aspects of digital security and privacy. \changes{The following sections} outline and explain \changes{only} the questions we used for this paper. The complete version of the questionnaire can be found in~\autoref{appendix:onlinesurvey}.

\subsubsection{Introduction} 
At the beginning of our questionnaire, we introduced the topic, our research interest and provided information on data handling and privacy. All participants gave informed consent before proceeding. 
\subsubsection{Demographics and Internet Usage}
The first part of our questions consists of demographic questions and questions concerning the general Internet usage of participants. Participants were asked what devices they use (Q1, Q2), whether they had been affected by different cybercrimes like malware (Q7), and whether and where they look for information about digital security (Q8). We based question Q7 about cybercrimes on a survey from the BSI (German Federal Office for Information security)~\cite{zindler-20-digitalbarometer}. 

\subsubsection{Misconceptions} \label{method:Misconceptions}
The misconceptions outlined in~\autoref{Misconceptions} were randomly presented grouped by topic to avoid sequencing effects.
Within each topic, the misconception and true statements were displayed in mixed and random order. For each statement participants were asked to indicate their agreement on a five-point rating scale by Rohrmann~\cite{rohrmann-07-verbal}, ranging from ``1--fully disagree'' ``to 5--fully agree.'' Additionally, the option ``I don't understand the statement'' was \changes{available}. 
\changes{For each topic, we formed a single misconception score, for which higher values indicate agreement \changes{with} misconceptions. 
Thereby, ratings for the correct statements were inverted.} 

\subsubsection{Concerns and Attitudes}\label{method:concernsandattitudes}
This questionnaire section started by asking participants how important it is to protect themselves from different threats like malware (Q17). We based this measurement on the survey by Story \etal \cite{story-21-web-privacy} and used a five-point rating scale by Rohrmann~\cite{rohrmann-07-verbal} ranging from ``1--not important'' to ``5--very important.'' Again, the option ``I don't understand the questions'' was \changes{available}. 

The next question (Q18) consisted of statements starting with ``How concerned are you...'' and covered concerns related to the misconceptions. An example statement is ``How concerned are you that when using messenger services your messages could also be read by unauthorized persons?'' Again we used a five-point rating scale raining from ``1--not concerned'' to ``5--very concerned'' and participants were able to answer ``I don't understand the question.''

Question Q19 focused on common attitudes which we grouped into three categories: 
\begin{enumerate}
\item Nobody is interested in my data (\eg, ``I am not rich or famous, so nobody is interested in accessing my data.'')
\item Encryption (\eg, ``Encryption is bad because it is used by hackers and criminals, \eg, for illegal activities.'')
\item Digital security is complicated (\eg, ``Digital security is annoying.''). 
\end{enumerate}

Participants were also asked which measures they use for their digital security, like updates~(Q20). We based the queried measures on commonly known security measures and expert advice~\cite{reeder-17-152-simple-steps}. 
We additionally listed a number of different data, \eg, name, address, health data and asked participants to indicate how important it is for them to protect the respective data on the Internet (\eg, from external access and theft-Q21). \changes{The response} options were again a five-point rating scale that ranged from ``1--not important'' to ``5--very important''.

The next question (Q22) asked participants how likely they \changes{believed} different groups or \changes{individuals were to} pose a risk to their digital security, \eg, unauthorized access to their personal data, stalk them online or restrict their access to digital services. We listed groups like \textit{family and friends}, \textit{work colleagues}, and \textit{officials from [insert country name], such as police, secret services, and the government}.
The response scale was a five-point scale ranging from ``1--not likely'' to ``5--very likely''.
Finally, we asked whether participants had practical experience in the computer science, computer technology, or information technology fields. Answer options were ``yes'', ``no'', or ``prefer not to answer''. 

The questionnaire contained \changes{a few} other questions that will not be discussed further here because they do not fit the research questions of this study (\eg, participants' communication behavior). The full survey is included in ~\autoref{appendix:onlinesurvey}.

\subsection{Survey Implementation and Panels}
\label{surveyimplementation}
To research and understand how experiences, concerns, and misconceptions about security-relevant issues differ around the world, we decided to conduct our representative survey in twelve countries on \changes{four} continents:
China~(CHN), Germany~(DEU), India~(IND), Israel~(ISR), Italy~(ITA), Mexico~(MEX), Poland~(POL), Saudi Arabia~(SAU), South Africa~(ZAF), Sweden~(SWE), Great Britain~(GBR), and the United States~(USA).
This list attempts to strike a balance between a wide geographic -- and to some extent cultural\footnote{We follow the approach of prior work and use country as a proxy for culture~\cite{sawaya-17-behavior, ur-13-cross-cultural, wang-11-concerns}.} -- diversity and countries \changes{amenable} to \changes{high-quality} online surveys. 
\changes{
More specifically, the following criteria led us to include the respective countries in our study:
China and India are the most populous countries in the world and likewise the countries with the most Internet users~\cite{setti2019analysis}.
In China in particular, Internet usage in terms of apps and providers differs from other countries, where services from US technology companies often dominate.
The populations of Germany, the United Kingdom and the United States are regularly the subject of studies due to their research and university landscapes and are correspondingly well researched.
The data collected here are therefore particularly well suited as a baseline and for comparisons with other studies.
We included Israel in our study because of its strong cyber security industry and education~\cite{adamsky2017israeli}.
Italy, Poland and Sweden are representatives of (Southern, Eastern and Northern) Europe, which is the geographical focus of our study. 
In addition, the Swedish population is considered to be particularly privacy-conscious~\cite{friedman2008personlig}.
Lastly, we selected Mexico, Saudi Arabia, and South Africa as populous representatives of Latin America, the Middle East, and Africa.
}
All countries surveyed together account for 42\% of the worlds' population. 

We first created a German-language preliminary version of our survey to estimate the processing time and ensure comprehensibility.
Based on feedback from a snowball sample of friends, families, and other researchers, we continuously improved the questionnaire. 
We also conducted a pilot study to test our questionnaire with 100 participants recruited via Prolific. We changed the wording of some statements and improved the questionnaire according to the results of the pilot testing. 

We commissioned Kantar Lightspeed, a full-service provider of online surveys that maintains panels world-wide, to conduct our survey including survey implementation and translation, participant recruitment, compensation, and data quality assurance.
In the first step, Kantar implemented the German-language version of the survey according to our requirements.
Next, a professional interpreter translated the survey into English, and several members of our research team carefully reviewed the translation.
The full English-language survey, including all \changes{implementation} instructions, can be found in~\autoref{appendix:onlinesurvey}.
Based on the English survey version, the translations into Arabic, Chinese, Hebrew, Italian, Polish, Spanish, and Swedish were then likewise carried out by professional translators commissioned by Kantar.

International surveys \changes{pose} a number of challenges \changes{related to} the required translations, \eg, technical terms or different educational systems. 
We have mitigated these problems by using professional translations, back translation with native speakers (see~\cite{sawaya-17-behavior, harbach-16-keep-on}), and the use of internationally established methods for measuring education, such as the ISCED~\cite{unesco-12-isced}.
\changes{
For the back translations, we recruited native speakers from our personal and professional circles to read through the survey with a participating researcher and back-translate it into English or German.
In the process, all translations proved to be of high quality, so that overall only a handful of translation errors had to be corrected.
}

Data collection in all twelve countries took place between mid December 2021 and early February 2022. Participants were chosen as a quota-representative sample in terms of \textit{age}, \textit{gender}, \textit{education}, \textit{region}, and, in the US, \textit{ethnicity}. Quotas were set by Kantar based on the most recent census data available.
\changes{
Kantar did not disclose the actual participant compensation to us.
However, they calculated with costs of \euro{2.51} (in China, India, Italy, Mexico, Poland, South Africa, UK and USA), \euro{2.61} (in Germany), \euro{3.20} (in Sweden), \euro{3.45} (in Saudi Arabia), and \euro{5.25} (in Israel) per completee. 
The compensation to be expected is -- at least in some of the countries -- below the respective legal minimum wage.
Rather, participant compensation appears to be in a similar range to compensation on crowdworking platforms~\cite{pater2021standardizing}.
However, we had no influence on the compensation and according to Kantar these amounts are in line with industry standards.
We cannot verify this assertion because we do not have comparative data on compensation for online panelists.
}

\subsection{Quality Assurance and Representativity}
The panel provider ensured data quality by removing speeders and participants who clicked certain answer patterns. \changes{Speeders were defined as participants answering the survey in less than $50\%$ of the median answer time}. To further increase the quality of the data, we included an attention check question (\textit{``This is a control question. Please click on the answer `mainly agree'.''}) in Q13 and participants who answered this question incorrectly were sorted out. 

\changes{
Kantar provided us with one data set per country.
We checked all data sets for complete or partial duplicate entries, word identical answers in open-ended question Q5 or click patterns, but could not detect any anomalies.
We then merged the country-specific data sets into a single final data set.
}

Representativity quotas for age and gender were matched with a maximum discrepancy of 4\% for all countries. 
The quotas for educational representativeness could not be met for China, India, Italy, Mexico, Saudi Arabia, and South Africa because their proportion of the population with a low level of education (ISCED levels 0-2~\cite{unesco-12-isced}) is relatively high and it is particularly difficult to reach them via (online) surveys.
Region quotas were met with high accuracy for all countries except Israel and Saudi Arabia, for which reliable data were not available in our panel provider's database.
In these two cases, we set and achieved the quotas for the regions using a best-effort approach based on publicly available data.

\subsection{Data Analysis}\label{method:data_analysis}
\changes{Before starting our analysis, we assessed the ``I don't understand the statement/question'' answers per item. For the misconception items (Q9 - Q16) the average frequency of this answer was $4.3\%$, which is rather low. We thus 
did not exclude any items based on this assessment. 
For the subsequent analysis the ``I don't understand this statement/question'' answers were excluded.} We then started with a descriptive analysis of the misconceptions statements. 
We calculated mean values and standard deviation for each misconception statement. 
We also combined the ratings of all misconception statements of one topic, \eg, E2EE~(Q9), to one mean value, \ie, one single score per misconception topic. 
Internal consistency for these scores was satisfactory~\cite{bland-97-cronbach} with \changes{all Cronbach's Alpha values above $0.70$, except for E2EE~(Q9)}, see~\autoref{tab:modelreliability}. 

For analyzing factors influencing the different misconceptions, we used the aforementioned misconception scores (Q9--Q16) for each topic as the outcome variable of our analysis. 
As our model included metric as well as continuous predictors, we used \changes{a special form of regression analysis -- namely} covariance analysis (lm model in R) \changes{-- as suggested in the literature}~\cite{luhmann-15-statistic}. For all metric predictor scales consisting of all the sub-questions were calculated. Internal consistency was acceptable~\cite{bland-97-cronbach} to good for these scales with \changes{all Cronbach's Alpha values above $0.80$, with only one exception (Q19--encryption; $0.51$)}. For an overview of all values, see~\autoref{appendix:predictorreliability} in \autoref{appendix:reliabilitytab}. We standardized these scales (\ie, all metric values), and the outcome variable for the subsequent analysis. Predictors and corresponding baselines are listed in~\autoref{tab:misconceptionpredictors} and are explained in~\autoref{Questionnairedesign}. 
For some predicting factors, we grouped several answer options into categories for our analysis: 
\begin{itemize}
    \item Q1. Device Usage--We grouped answers in four categories, no device usage (\textit{baseline}), using one of the named devices (\textit{few}), using two to three devices (\textit{moderate}), using four to six devices (\textit{many}). 
    \item Q7. Experiences--We grouped the answers into two categories, participants with no experiences with cybercrime (\textit{no--baseline}) and participants having experiences with cybercrime (\textit{yes}).
    \item Q8. Information--We grouped the answers into two categories, participants not seeking information about digital  security (\textit{no--baseline}) and participants looking for this information (\textit{yes}).
    \item Q20. Measures taken--We grouped the measures taken by participants to secure their devices and accounts similar to Q1 in four categories, none (\textit{baseline}), one to five measures (\textit{few}), six to nine measures (\textit{moderate}), ten to thirteen measures (\textit{many}).
    \item Q25. IT Experience--We grouped the answers into two categories, participants being experienced with IT security or related fields (\textit{yes}) and participants without experience in this field (\textit{no--baseline}).
\end{itemize}
\input{data/tab-model-reliability}

For the analysis of influential factors for misconceptions, we conducted one covariance analysis per misconception topic (Q9--Q16). We started each covariance analysis with only country as predictor and iteratively included predictors based on their contribution to the model, i.e., their coefficient of determination ($R^2$), starting with the highest one. We included the predictors to the model iteratively based on three model fit criteria: Maximal coefficient of determination ($R^2$), minimal Akaike information criterion (AIC) and ANOVA test (between the two models, with and without the new predictor). The model resulting of these iterations was afterwards compared to a model including all the predictors, using the same criteria. We report the model with the best fit for each misconception topic (Q9--Q16) in~\autoref{tab:misconceptionpredictors}. Therefore, the different models do not consist of the same predictors. Which predictors were excluded, is listed for every model respectively in~\autoref{appendix:modelVariables}. We used standardization for predictors and outcome variable to compare results across models, thus all reported estimates are standardized. 

Results of the covariance models are shown in~\autoref{tab:misconceptionpredictors}. Positive estimates indicate positive influences on having misconceptions, negative estimates indicate a negative influence on holding misconceptions.

\subsection{Ethics and Data Protection}
Our institution does not have an institutional review board~(IRB) nor an ethics review board~(ERB) that we could consult for our study. 
Nonetheless, we followed best practices of user research~\cite{dhs-12-menlo-report} and data protection guidelines, including the European GDPR.
All data protection measures were reviewed and approved by our institution's data protection office.
In addition, Kantar, our panel provider, has committed to abide by the ICC/ESOMAR Code of Conduct, which sets out ethical and professional obligations when conducting (online) surveys~\cite{icc-07-code}. 
The panel provider signed an agreement with our institution to \changes{comply with strict} GDPR guidelines \changes{for participants in all countries surveyed}. 
We also provided a debriefing document stating reasons why and which of the statements in our survey were true and which represented misconceptions. 
Due to technical reasons, the panel provider emailed the debriefing to the participants after the survey. 

\subsection{Limitations}
\changes{
Although we have done our best to include one country for each world region, our country sample is primarily focused on the Eurasian continent.
This is particularly related to the availability of high-quality online panels.
It is possible that the inclusion of additional countries from Africa, Asia, and Latin America could provide further insight into privacy and security perceptions and behaviors in these areas.
}
For the same reason, we also lack data on different ethnicities for most countries. 
However, we were able to include representative quotas on ethnicity for the US sample. 
In addition, it is very difficult to reach older people and people with \changes{little} education with online panels in general (see~\cite{redmiles-19-mturk-generalization, tang-22-generalize}) and especially in countries in the global South. 
Therefore, we could not meet the representative education quotas for a number of countries. 
While we believe that we have reached and studied a broadly representative sample of Internet users in these countries as well, it may be worthwhile to specifically address the digital security needs of people with lower levels of education again in future studies.
\changes{
Two of the scales included in our survey exhibited less than acceptable internal consistencies ($\alpha$<$0.70$, Q19-E2EE and Q9-E2EE). 
We can only speculate that this may be due to the rather complex, unfamiliar subject matter of these scales for the participants. 
Because all items were rated as important in our pilot test, we refrained from excluding these scales. 
However, the lower internal consistencies must be taken into account when drawing conclusions from the corresponding results.
}

%% file: data/tab-model-reliability.tex
\begin{table}
    \centering
    \small
    \Description[CovarianceAnalysis]{This table has 8 rows. Every row shows a misconception and the matching coefficient of determination (i.e., explained variance ($R^2$)) and the reliability coefficient Cronbach's alpha ($\alpha$) together with the number of participants (n) that were included in respective the covariance model.}
    \caption{Misconception scales and covariance model criteria per topic for Q9--16 including number of participants (N), misconception \changes{topic}, \changes{adjusted} coefficient of determination (explained variance; \changes{adjusted} $R^2$), and reliability coefficient Cronbach's alpha ($\alpha$).}
    \label{tab:modelreliability}
    \vspace{-0.3cm}
\begin{tabular}{@{}clcc@{}}
    \toprule
    N & Misconception Topic & \changes {adjusted $R^2$} & $\alpha$ \\
    \midrule
    11484 & Q9. E2EE Messenger & 0.14 & 0.64 \\
    11476 & Q10. HTTPS & 0.13 & 0.76 \\
    11558 & Q11. Wi-Fi & 0.10 & 0.82 \\
    11001 & Q12. VPN & 0.08 & 0.79 \\
    11641 & Q13. Password and Login & 0.38 & 0.77 \\
    11621 & Q14. Device Security & 0.30 & 0.73 \\
    11626 & Q15. Malware and Deception & 0.40 & 0.81 \\
    11391 & Q16. Private Browsing Mode & 0.16 & 0.94 \\
    \bottomrule
\end{tabular}
\end{table}

%% file: sections/06_results.tex
\section{Results}
\label{sec:results}
In this section, we first describe our sample based on demographic data and device usage. 
We then briefly compare misconception prevalence per country and finally take a look at the factors that significantly influence misconceptions on the eight different security and privacy relevant topics (see~\autoref{Misconceptions}). 

\subsection{Sample Description}
In \autoref{tab:demographics}, we provide demographic information (gender, age, and education) about our participants as well as information on the used devices \changes{and median completion times} \changes{per} country. 
In our sample, smartphones, laptops and PCs as well as tablets were the most used devices with rather similar usage rates across countries. 
Smart speakers and wearables were much less used, with a tendency of higher usage rates in Asian countries.

\input{data/tab-demographics}

\subsection{\changes{RQ1: Misconceptions Around the World}}
The descriptive analysis of our \changes{misconceptions} \changes{-- to answer RQ1--} revealed, that participants around the world are rather unsure about the queried misconceptions. The mean values for all misconception scores (\changes{see~\autoref{method:Misconceptions}}) are located around the middle of the scale (``3 -- neutral''). However, in the following sections we describe differences between countries and misconception topics and highlight specific misconceptions participants mostly agreed or disagreed to ($mean > 4$ and $mean < 2$).

\subsubsection{Misconception Scores (Q9--Q16)}
\begin{figure}
	\centering
	\includegraphics[width=0.8\textwidth]{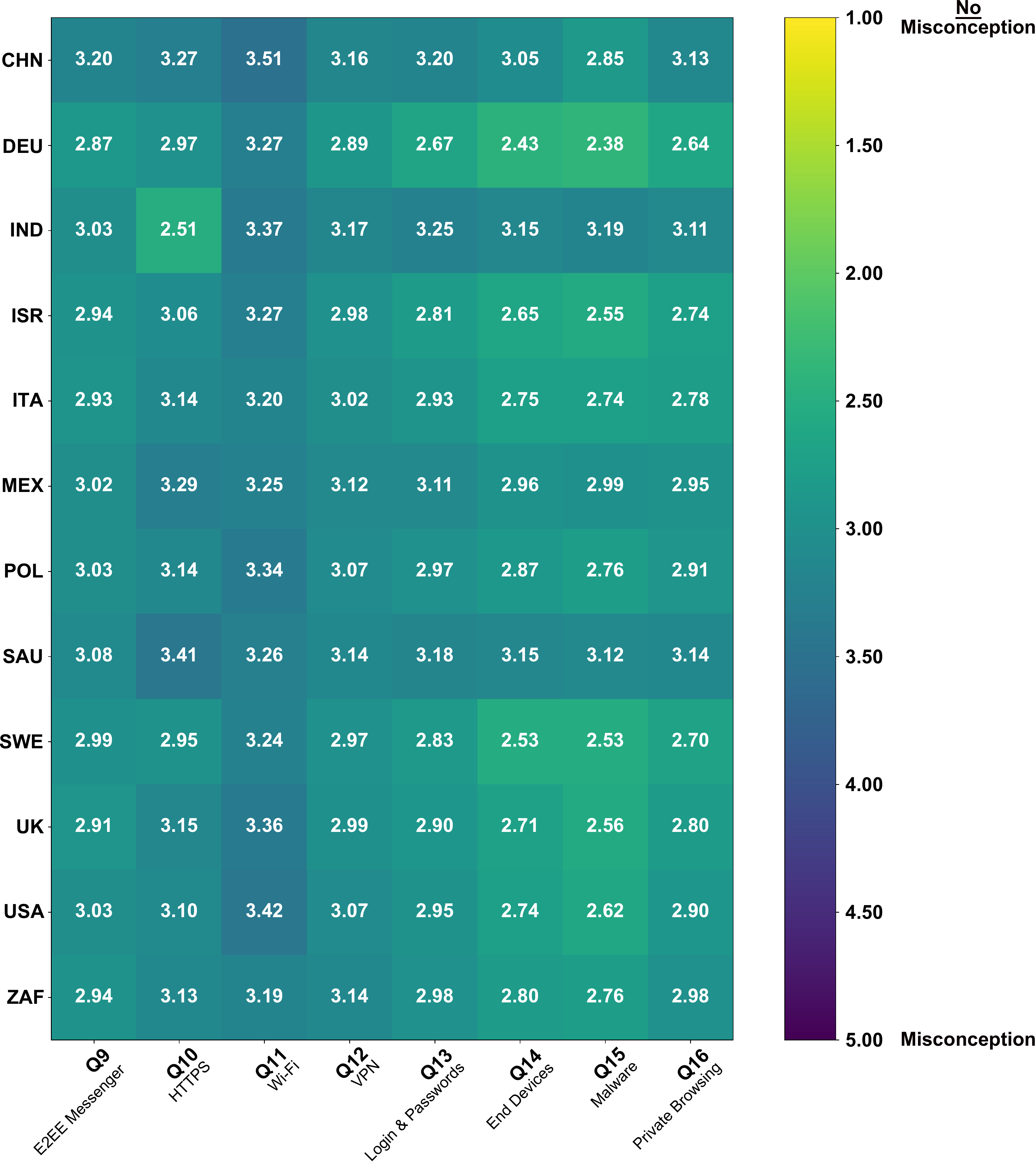}	
	\Description[MisconceptionHeatMap]{In a heatmap the mean values of the 8 misconceptions we analyzed are displayed for all 12 countries. The heatmap is organized in cells (in total there are 12*8=96 cells). Every cell holds the mean value, and a color visualizes the the mean value. Higher mean values are displayed in a darker color than lower mean values.}
	\caption{Comparison of mean values per country across our eight misconception topics. Darker colors show agreement with misconceptions and lighter colors show disagreement with misconceptions.}
	\label{fig:overallheatmap}
\end{figure}

For all misconceptions we found moderate agreement, ranging around the middle of the scale with only very slight outliers. \autoref{fig:overallheatmap} illustrates the score mean values for each misconception topic in all surveyed countries. For these score mean values, values closer to five show agreement \changes{with} misconceptions and mean values closer to one indicate disagreement \changes{with} the respective misconception topic. 

Across all countries and topics, we observed score mean values from $M = 2.51$ up to $M = 3.51$ with standard deviations ranging from $SD = 0.29$ up to $SD = 0.69$. 
\changes{These results indicate} 
that participants were more or less unsure about a lot of the posed misconceptions. The rather small standard deviations show that our data is gathered around the mean, hinting at a rather small amount of variation in the participants' answers. This shows that participants \changes{from the same} country rated the statements similar. 
German participants showed the least agreement \changes{with} nearly all misconception topics, indicating that they least believed the misconceptions, even though most of the values were around the middle of the scale. Participants from China and India indicated the highest agreement \changes{with} many misconceptions across topics, with some mean values, \eg, Q11 in China, leaning towards agreement (``4 -- mainly agree''). 
We observed the smallest mean value for German participants on the topic of malware~(Q15), with a score mean of $M = 2.38$ and a standard deviation of $SD = 0.56$. 
We found the highest score mean with $M = 3.51$ for Chinese participants and misconceptions relation for Wi-FI~(Q11). 
The misconceptions for Wi-Fi~(Q11) got the most agreement across all countries and misconceptions about malware~(Q15) got the most disagreement. 

\subsubsection{Dominant Misconception Statements}
When looking at specific misconception statements, we found participants' agreement ($M > 4$) or disagreement ($M < 2$) to thirteen statements compromising misconceptions as well as correct statements of digital security and privacy tools or concepts. 

We closer investigated statements with mean values less than two and greater than four, \changes{indication} clear disagreement and agreement \changes{with those (mis)conceptions}. 

One misconception that participants from all countries except Saudi Arabia ($M = 3.96$) agreed with ($M > 4$) was the importance of changing passwords regularly (Q13-6): ``It is important for the security of my user accounts to regularly change the password.'' This was an advice given to users for many years, but regularly changing the password only puts a burden on users and does rather not improve security~\cite{habib-18-expiration, mayer-17-pw-misconceptions, chiasson-15-pw-expiration}, except for when the account is compromised. We see that users still believed that this advice is true even though it is no longer given but rather discouraged. Another misconception participants from all countries agreed to was ``My PC can get infected with malware by clicking on a link''~(Q15-10) -- which is only true in cases of sophisticated zero-click attacks like Pegasus~\cite{marczak-18-pegasus} that only aim at single high-value targets. \changes{In the vast majority of} cases, when browser and operating system are kept up-to-date, clicking on a link is not sufficient to install malware on a computer. Only the download and further interaction with a file would be dangerous. 
Participants from India agreed ($M = 4.05$) to the misconception that HTTPS indicates a websites' trustworthiness~(Q10-4), when in fact HTTPS only indicates a secure connection. Participants from other countries (except for Israel, Germany, and the US) also rather agreed to this statement ($M > 3.5$). We rated this statement as a misconception as even fraudulent websites can set up HTTPS and thus, the user transmits their data over a secure connection to the offenders.

We observed agreement ($M = 4.09$) from Chinese participants for the misconception statement ``When I am connected to a public Wi-Fi, it is easy to infect my device with malware''~(Q11-3). We rated this statement as a misconception, as it does not consider the device configuration, like up-to-date anti-malware components, which will protect from being infected with malware. For this statement all other countries also tended to agree with mean values above $M = 3.5$. 

We observed a misconception regarding two factor authentication for participants from Saudi Arabia and Germany, who agreed ($M = 4.03$, $M = 4.10$) to the statement ``I have to log in to online banking with two processes so that the connection is encrypted, for example, with a password and TAN (transaction number)''~(Q13-13). This shows \changes{participants misunderstand and confuse} encryption and authentication, \changes{which} was also found by Krombholz \etal~\cite{krombholz-19-mental-models-https}. 
Participants from Saudi Arabia ($M = 4.03$) as well as from China ($M = 4.3$) believed that the content of a website reveals potential threats emerging from this website (``Is it more likely to pick up malware from visiting a porn website than visiting a website on the topic of sport''~Q15-3). \\

We also observed disagreement \changes{with} some misconception statements. Participants from South Africa disagreed ($M = 1.83$) to the statement that locking ones device is not necessary~(Q14-6), indicating that they might think it is necessary security-wise, which is true. German participants disagreed ($M = 1.74$) to the statement ``I can click on attached files without concern for an email that is addressed to me directly.''~(15-11), revealing that they were familiar with phishing and the fact that phishing emails can be directly addressed to the recipient. We observed disagreement ($M = 1,87$) from German participants to the statement ``As long as a website looks official, I can enter my login data without concern''~(Q15-12). As malicious websites often imitate real websites to phish people, the look and feel of a website is not a sufficient indication for a real or fake evaluation.

Similar to disagreeing to misconception statements we also \changes{observed} agreement \changes{with} true statements ($M > 4$). Participants from all countries agreed (means ranging from $4.19$ to $4.51$) to the statement that special characters and numbers lead to increased password security. We rated this statement as true, as generally speaking, the security of a password is enhanced when the number of possible combinations is increased by using additional digits like numbers or special characters. 
Shoulder-surfing is a security risk participants from all countries were aware of, with highest awareness (agreement values for Q14-1) in Germany, Poland, Sweden, UK, the US, and South Africa. The possibility for unnoticed malware on ones' device was also \changes{familiar to} all participants with highest awareness in Germany and Sweden (agreement \changes{with} Q15-7). The concept of ransomware was somewhat known by all participants (mean values for all countries $> 3.5$) with highest agreement \changes{values in} Germany, Israel, and Sweden~(Q15-8). We observed the same for the concept of phishing~(Q15-16), with mean values for all countries ranging between $4$~(China) and $4.4$~(Germany).

\subsection{\changes{RQ2: Factors Predicting Misconceptions}}

In \changes{this section,} \changes{we report on our results regarding RQ2, showing which} factors predict security-related misconceptions. 
Even though we calculated the covariance models per misconception topic, we ordered results by predicting factors for better comprehension. Due to standardization \changes{of the predictors} we were able to compare factors across models. 
All significant predictors with estimates and corresponding significant levels are listed in \autoref{tab:misconceptionpredictors}. Overall, we observed the highest estimate values for country of residence followed by security measures taken, attitudes regarding privacy and security as well as device usage. 
The \changes{adjusted} $R^2$ values for every misconception topic are shown in \autoref{tab:modelreliability}. \changes{Adjusted} $R^2$ represents the proportion of variance for the outcome variable, that is explained by the predictors \changes{(considering the number of predictors)}. We observed mixed results. For misconceptions regarding passwords and login processes~(Q13), device security~(Q14) as well as malware and deception~(Q15), our prediction factors explained 30\% -- 40\% of variance. For all other topics, however, our predictors only accounted for 8\%-16\% of the variance. 

\input{data/tab-misconception-predictors}

\subsubsection{\changes{Country of Residence Predicts Belief in Misconceptions}}
Country of residency proved to be the best predictor for the studied misconceptions -- indicated by the largest significant estimates (except for Wi-Fi~Q11), which showed participants had more (positive estimates) or less (negative estimates) misconceptions compared to participants from Germany. 

We observed that Western and non-Western countries differed especially in magnitude of estimates. We found the largest estimates and thus greatest differences \changes{compared} to Germany, for India, Saudi Arabia, and China. 

For almost all misconception topics, except those related to Wi-Fi, the estimates were highest for either Chinese or Indian participants. For misconceptions related to Wi-Fi the highest estimate existed in South Africa. 
Chinese and US participants were more likely to believe in all misconceptions (positive significant estimates for all topics) than German participants. The same applied to the participants from India, Poland, and Sweden, who were significantly more likely to agree \changes{with} not all but most of the misconception topics, compared to German participants. 
For all remaining countries at least one estimate was negative, showing that participants were less likely to believe (certain) misconceptions than German participants. 
We observed the smallest discrepancy between holding misconceptions for Israeli and German participants (estimates range from $-0.12$ to $0.23$). 
We generally found higher estimates for non-Western countries (China, India, Mexico, Saudi Arabia, South Africa) compared to Western countries. 
Compared to German participants, participants across all other countries were more likely to believe in misconceptions related to malware~(Q15), device security~(Q14), and passwords~(Q13) indicated by higher positive estimates. For the predictor \textit{country} we observed the lowest estimates for misconceptions related to Wi-Fi~(Q11), HTTPS~(Q10), and E2EE~(Q9). For these topics the least differences existed between Germany and the other countries.

\subsubsection{\changes{Demographics are Rather Small but Significant Predictors for Misconceptions}}
For the demographic predictors age, gender, and education, we observed mixed results with age as a significant predictor for most cases (except for HTTPS) and gender as a significant predictor for only five misconception topics (E2EE, Wi-Fi, VPN, passwords and private browsing). Also the estimates for age were larger compared to gender and education. 

Compared to younger participants, participants older than 25 were generally more likely to believe in misconceptions, with slightly larger estimates for older participants than for those between $25-39$ years. Except for the topics malware and HTTPS, participants older than 25 were more likely to hold misconceptions than participants younger than 25. We observed highest estimates for the topics Wi-Fi and private browsing, with the biggest differences between very young (18-24) and older participants -- the highest value was observed for participants 55+ ($0.27$). Participants older than 40 were less likely to believe in misconceptions regarding malware, compared to the young baseline. 

Compared to men, women were more prone to hold misconceptions about E2EE, VPN, passwords, and private browsing. Misconceptions regarding Wi-Fi were found more frequently with men than women. Effects in all cases were -- however -- small.

Most estimate sizes for education were also rather small, with medium and low education as positive significant predictors for believing in misconceptions across topics, with an exception for Wi-Fi. We found no significant differences in believing misconceptions regarding end-to-end encryption and HTTPS between different levels of education. Participant with less than high education were less likely to believe in misconceptions regarding Wi-Fi. For the other misconceptions (VPN, passwords, device security, malware, private browsing), less than high education was associated with believing more in misconceptions.

For misconception statements related to HTTPS none of the demographic factors \changes{were} significant predictors. 

As shown in~\autoref{tab:demographics}, we also considered device usage a demographic factors. The data showed only a few significant, but rather large estimates for the predictor device usage. Participants who used more than two of the listed devices were more likely to believe in misconceptions related to Wi-Fi than those who used none of the queried devices. Participants who used more than four devices were also more likely to believe in misconceptions regarding E2EE and VPN. 

\subsubsection{\changes{Experience with Cybercrime Predicts Disbelief in Misconceptions}}
Our questionnaire assessed participants' experiences with cybercrime (Q7) as well as their professional IT experience (Q25). Our data showed a positive association of prior cybercrime experience with believing less in misconceptions, whereas professional IT experience was a positive predictor for misconceptions. 
Prior experience with some sort of cybercrime (all participants indicating experience with at least one type of crime mentioned in Q7), was significantly associated with believing less in misconceptions about HTTPS, VPN, device security, malware, and private browsing. However, the estimate values were rather small ($< 0.1$). 
Contrary to this, we found that prior professional experience with IT predicted believing in misconceptions regarding HTTPS, VPN, passwords, device security, and malware, also with small estimates ($< 0.1$). 
Familiarity with security or privacy (Q7, Q25) does not predict misconceptions regarding the topics end-to-end encryption and Wi-Fi. 

\subsubsection{\changes{Protection Devices and Data Predicts Disbelief in Misconceptions}}
Questions Q17 and Q21 in our questionnaire both asked how important participants consider protecting their devices and data, \eg, from malware~(Q17), and how important it is to them to protect specific data types, \eg, private photos. We found participants, who generally think it is important to protect their devices and data, were less likely to hold misconceptions regarding end-to-end encryption, passwords, device security, malware and private browsing. 
Participants who found it rather important to secure specific data types online were more likely to believe misconceptions of all topics, except for Wi-Fi. Despite their significance, both estimate values were rather small ($< 0.1$). 
Misconceptions regarding Wi-Fi were not predicted by protection importance as estimates for both question Q17 and Q21 are not significant. 

\subsubsection{\changes{Using Countermeasures Predicts Disbelief in Misconceptions}}
Taking active measures for more privacy and security, like using end-to-end encryption, will increase users' digital security and privacy but seeking information on these topics might also do so. Surprisingly, we found that participants who actively seek information on digital security were more likely to believe in misconceptions regarding all topics (except for Wi-Fi), than participants who did not look for this kind of information (baseline). Estimates were \changes{slightly} higher than those for the aforementioned predictors, ranging from $0.08$ (passwords~Q13) to $0.13$ (HTTPS~Q10). 
However, we observed that using measures to stay safe online was a negative predictor for believing in most of the queried misconceptions, thus participants who took measures were less likely to believe misconceptions than those who did not take any security or privacy measures (baseline). These estimates were in the mid range compared to all estimate absolute values, ranging from $-0.09$ to $-0.38$. Most differences in believing in misconceptions existed between participants who did not take any security measures and those who took at least a moderate amount ($5$ or more) of protection measures, with highest negative estimates for the topics passwords~(Q13) and malware~(Q15). However on the contrary, participants taking moderate or many protection measures were more likely to hold misconceptions related to Wi-Fi~(Q11) than those who took no such measures. 
Taking security and privacy protection measures was \changes{not a} predictor for \changes{holding} misconceptions regarding end-to-end encryption~(Q9) and HTTPS~(Q10).  

\subsubsection{\changes{Thinking Digital Security is Complicated Predicts Belief in Misconceptions}}
The questionnaire also included questions on attitudes towards using digital security, that we grouped into three scales, see~\autoref{method:concernsandattitudes} for details. 
Participants who agreed \changes{with} more of the attitude statements, were also more likely to believe in misconceptions, regarding all misconception topics, except for the combination of participants having more attitudes related to E2EE and the misconceptions topic HTTPS. 
Participants who held attitudes like, \eg, end-to-end encryption is only for paranoid people and has more disadvantages than advantages were more likely to hold misconceptions across most topics. However, these effects were rather small, with estimates ranging from $0.03$ to $0.12$. Participants with these attitudes were less likely to believe misconceptions regarding HTTPS. 
We observed the highest estimates for attitudes related to third-party-interest in ones data. Participants who did not believe anybody was interested in their data and thus did not consider themselves at risk were more likely to believe misconceptions related to all topics, except for Wi-Fi. 
Participants who thought securing data and profiles was complicated, also tended to believe most misconceptions, but with smaller effects compared to the aforementioned attitudes (estimates $< 0.1$). 

\subsubsection{\changes{Being More Concerned Predicts Belief in Misconceptions}}
We also queried participants about the amount of concern for different threats like data theft (Q18) and who they considered a risk to their security (Q22). We found that for both, participants who were more concerned and those who thought more groups pose a risk, to be more likely believing in almost all misconceptions. 
Participants who were more concerned (higher mean value Q18), were more likely to believe misconceptions regarding all topics, except for device security. Compared to other prediction factors, however, the estimate values were rather low ($<= 0.1$). 
Similarly, participants who viewed more groups of people (like hackers and companies) as risks for their digital security more likely believed misconceptions related to all queried topics~(Q9-Q16), but only with slightly higher estimates ($<= 0.14$).

%% file: data/tab-demographics.tex
\begin{center}
    \addtolength{\leftskip} {-2cm} 
    \addtolength{\rightskip}{-2cm}
\begin{table*}[htbp]
\small
\Description[Demographics]{The table shows the relative percentage of gender, age and education for the 12 countries we performed our survey in. Additionally, the absolute percentage of device usage for devices like smartphones, tablet, smart speakers are displayed.}
\centering
\caption[Participant Demographics]{\label{tab:demographics}
Participant demographics. Data for gender, age, education, \changes{and completion time in minutes} as delivered by our panel provider.
Information about participants' device use was collected in the questionnaire. Ethnicity was collected only for the US by our panel provider (White: 70.3\%, African American: 11.5\%, Hispanic/Latino: 9.6\%, Asian: 6.0\%, Other: 2.2\%).}
\begin{tabular*}{\textwidth}{@{}l@{\extracolsep{\fill}}*{12}{S[table-format=2.1]}@{}}
\toprule
& \multicolumn{12}{c}{\textbf{Country}}\\

\cmidrule{2-13}
&   \multicolumn{1}{c}{\textbf{CHN}} & \textbf{DEU} & \textbf{GBR} & \textbf{IND} & \textbf{ISR} & \textbf{ITA} & \textbf{MEX} & \textbf{POL} & \textbf{SAU} & \textbf{SWE} &  \textbf{USA} & \multicolumn{1}{c}{\textbf{ZAF}} \\


&  \multicolumn{1}{c}{(1025)} & \multicolumn{1}{c}{(1019)} & \multicolumn{1}{c}{(1018)} & \multicolumn{1}{c}{(1018)} & \multicolumn{1}{c}{(1024)} & \multicolumn{1}{c}{(1019)} & \multicolumn{1}{c}{(1045)} & \multicolumn{1}{c}{(1054)} & \multicolumn{1}{c}{(1021)} & \multicolumn{1}{c}{(1049)} & \multicolumn{1}{c}{(1029)} & \multicolumn{1}{c}{(1048)}\\

\midrule
\textbf{Gender} & \textbf{\%} & \textbf{\%} & \textbf{\%} & \textbf{\%} & \textbf{\%} & \textbf{\%} & \textbf{\%} & \textbf{\%} & \textbf{\%} & \textbf{\%} & \textbf{\%} & \textbf{\%} \\

Female &  46.6 & 49.5 & 51.1 & 46.0 & 49.7 & 52.0 & 49.2 & 50.3 & 41.5 & 50.4 & 51.6 & 50.2 \\
Male &  51.9 & 49.2 & 48.3 & 50.9 & 44.5 & 46.7  & 47.0 & 44.5 & 49.8 & 48.6 & 46.8 & 44.6\\
Other &  1.5 & 1.3 & 0.6 & 3.1 & 5.8 & 1.3 & 3.8 & 5.2 & 8.7 & 1.0  & 1.6 & 5.2 \\

\midrule
\textbf{Age} & \textbf{\%} & \textbf{\%} & \textbf{\%} & \textbf{\%} & \textbf{\%} & \textbf{\%} & \textbf{\%} & \textbf{\%} & \textbf{\%} & \textbf{\%} & \textbf{\%} & \textbf{\%} \\

18--24 &  9.6 & 7.4 & 8.6 & 19.8 & 14.1 & 8.1 & 19.3 & 9.8 & 19.0 & 11.0 &  11.0 & 21.9 \\
25--39 & 35.1 & 22.7 & 25.5& 37.8 & 31.3 & 20.3 & 36.3 & 28.5 & 54.6 & 23.3 &   25.1 & 40.7 \\
40--54 & 41.8 & 27.4 & 26.7 & 25.7 & 24.3 & 29.6 & 26.9 & 24.7 & 24.0 & 25.3 &  27.0 & 23.8 \\
55+ &  13.5 & 42.5 & 39.2 & 16.7 & 30.3 & 42.0 & 17.5 & 37.0 & 2.4 & 40.4 &  36.9 & 13.6 \\

\midrule

\textbf{Education} & \textbf{\%} & \textbf{\%} & \textbf{\%} & \textbf{\%} & \textbf{\%} & \textbf{\%} & \textbf{\%} & \textbf{\%} & \textbf{\%} & \textbf{\%} & \textbf{\%} & \textbf{\%} \\
        
Low (ISCED 0-2)
& 8.0 & 15.4 & 18.8 & 3.8 & 9.2 & 15.5 & 30.6 & 3.4 & 7.5 & 9.7 &  2.6 & 25.5 \\
Medium (ISCED 3-4)
& 36.3 & 51.9 & 33.3 & 36.0 & 34.9 & 54.3 & 28.7 & 58.5 & 38.8 & 43.4 &  40.5 & 39.7 \\
High (ISCED 5-8)
& 55.4 & 32.4 & 47.7 & 58.0 & 54.3 & 29.9 & 40.1 & 37.8 & 53.1 & 45.6 & 55.6 & 31.6 \\
Other
& 0.3 & 0.3 & 0.2 &  2.2 & 1.6 & 0.3 & 0.6 & 0.3 & 0.6 & 1.3 & 1.3 & 3.2 \\


\midrule

\textbf{Q1. Device Use} & \textbf{\%} & \textbf{\%} & \textbf{\%} & \textbf{\%} & \textbf{\%} & \textbf{\%} & \textbf{\%} & \textbf{\%} & \textbf{\%} & \textbf{\%} & \textbf{\%} & \textbf{\%} \\
Smartphone
& 99.8 & 92.7 & 88.8 & 98.8 & 96.9 & 97.7 & 94.5 & 96.4 & 97.8 & 95.2 & 88.8 & 97.8 \\
Tablet
& 51.6 & 45.4 & 50.6 & 37.5 & 30.0 & 52.0 & 43.0 & 38.3 & 45.6 & 49.2 &  43.5 & 34.6 \\
Laptop
& 72.3 & 68.7 & 71.6 & 76.0 & 72.0 & 73.2 & 59.3 & 83.6 & 69.2 & 73.5 &  60.4 & 74.8 \\
Stationary PC
& 63.1 & 49.0 & 37.1 & 41.9 & 61.3 & 54.7 & 41.9 & 45.2 & 42.7 & 44.7 &  42.2 & 30.7 \\
Smart Speaker
& 36.7 & 17.8 & 26.8 & 36.1 & 7.1 & 25.9 & 22.7 & 6.2 & 17.7 & 12.5 & 24.5 & 8.1 \\
Wearable
& 32.2 & 14.1 & 20.1 & 38.3 & 16.5 & 23.2 & 15.5 & 25.0 & 34.9 & 13.7 & 16.9 & 18.4 \\



\midrule

\changes{\textbf{Completion time}} & min & min & min & min & min & min & min & min & min & min & min & min \\
Median & 19.5 & 21.5 & 19.7 & 24.8 & 24.0 & 22.2 & 30.2 & 25.9 & 24.6 & 24.2 & 21.9 & 32.3 \\

\bottomrule
\end{tabular*}
\end{table*}
\end{center}


%% file: data/tab-misconception-predictors.tex
\begin{center}
    \addtolength{\leftskip} {-2cm} 
    \addtolength{\rightskip}{-2cm}
\begin{table*}
\small
\Description[Misconception]{This rather complex table shows the results of the covariance analysis for the 8 misconceptions we investigated. The predictors are age, gender, education, device usage, experiences, information, prevention, concerns, attitude, security measures taken, data protection, potential attackers, professional IT experience and the country of residence. Only significant estimates are presented. The significance is visualized via stars.}
\centering
\caption[Misconceptions]{\label{tab:misconceptionpredictors}
Covariance Analysis per misconception topic. Only significant estimates (rounded to second decimal) are reported. Positive estimates indicate positive influences on having misconceptions, negative estimates indicate a negative influence on holding misconceptions. Significance levels are indicated with stars (*$p<.05$, **$p<.01$, ***$p<.001$). All estimates are standardized (see~\autoref{method:data_analysis}). Sample size, \changes{adjusted} R² and Cronbach's Alpha per topic are shown in \autoref{tab:modelreliability}.}
\vspace{-0.35cm}
\begin{tabular*}{\textwidth}{@{}l@{\extracolsep{\fill}}*{8}{S[table-format=3.2]}@{}}
\toprule
\textbf{Predictor} & \multicolumn{8}{c}{\textbf{Estimate}} \\
\cmidrule{2-9}
& \multicolumn{1}{c}{Q9} & \multicolumn{1}{c}{Q10} &  \multicolumn{1}{c}{Q11} & \multicolumn{1}{c}{Q12} & \multicolumn{1}{c}{Q13} & \multicolumn{1}{c}{Q14} & \multicolumn{1}{c}{Q15} & \multicolumn{1}{c}{Q16}\\
& \multicolumn{1}{c}{E2EE} & \multicolumn{1}{c}{HTTPS} & \multicolumn{1}{c}{Wi-Fi} & \multicolumn{1}{c}{VPN} & \multicolumn{1}{c}{Passwords} & \multicolumn{1}{c}{Device Security} & \multicolumn{1}{c}{Malware} & \multicolumn{1}{c}{Priv. Browsing}\\

\midrule\multicolumn{8}{@{}l@{}}{\emph{Age (baseline: 18-24)}} \\
25-39 & 0.07* & & 0.13*** & & 0.06* & & & 0.17*** \\
40-54 & & & 0.20*** & & & & -0.11*** & 0.19*** \\
55+ & 0.11*** & & 0.22*** & 0.10** & & 0.06* & -0.07** & 0.27*** \\

\midrule\multicolumn{8}{@{}l@{}}{\emph{Gender (baseline: Male) }}\\
Female & 0.07*** & & -0.09*** & 0.06*** & 0.04* &  &  & 0.16*** \\

\midrule\multicolumn{8}{@{}l@{}}{\emph{Education (baseline: High - ISCED 5-8)}}\\
Low (ISCED 0-2) & & & -0.20*** & & 0.07** & 0.09** & 0.19*** & 0.13*** \\
Medium (ISCED 3-4) & & & -0.05* & 0.06** & 0.04** & 0.04* & 0.06*** & 0.08*** \\

\midrule\multicolumn{8}{@{}l@{}}{\emph{Q1. Device Usage (baseline: none)}}\\
Few (1) & & & & & & & & \\
Moderate (2-3) & & & 0.50** & & & & & \\
Many (4-6) & 0.38* & & 0.60*** & 0.16* & & & & \\

\midrule\multicolumn{8}{@{}l@{}}{\emph{Q7. Experience (baseline: No)}}\\
Yes & & -0.07*** & & -0.08*** & & -0.07*** & -0.06*** & -0.03** \\
\midrule\multicolumn{8}{@{}l@{}}{\emph{Q8. Information (baseline: No)}}\\
Yes & 0.01*** & 0.13*** & & 0.11*** & 0.08*** & 0.11*** & 0.10*** & 0.09*** \\

\midrule
Q17. Prevention & -0.03** & & & & -0.07*** & -0.07*** & -0.07*** & -0.03** \\
Q18. Concerns & 0.10*** & 0.04** & 0.13*** & 0.09*** & 0.10*** & & 0.03*** & 0.06*** \\
\midrule\multicolumn{8}{@{}l@{}}{\emph{Q19. Attitudes}}\\
E2EE & 0.12*** & -0.09*** & & & 0.11*** & 0.02* & 0.05*** & 0.03* \\
Interest & 0.15*** & 0.21*** & & 0.12*** & 0.30*** & 0.36*** & 0.37*** & 0.21*** \\
Complicated & 0.04** & & 0.07*** & & 0.09*** & 0.05*** & 0.03*** & \\

\midrule\multicolumn{8}{@{}l@{}}{\emph{Q20. Measures Taken (baseline: None)}}\\
Few (1-5) & & & & & -0.16** & & -0.09*** & \\
Moderate (6-9) & & & 0.14** & -0.15* & -0.30*** & & -0.34*** & -0.17** \\
Many (10-13) & & & 0.29*** & -0.14* & -0.31*** & -0.14* & -0.38*** & -0.17* \\

\midrule
Q21. Data Protection & 0.03* & 0.06*** & & 0.04*** & 0.08*** & 0.11*** & 0.11*** & 0.08*** \\
Q22. Potential Attackers & 0.10*** & 0.07*** & 0.15*** & 0.05*** & 0.14*** & 0.04*** & 0.05*** & 0.05*** \\
\midrule\multicolumn{8}{@{}l@{}}{\emph{Q25. Professional IT Experience (baseline: No)}}\\
Yes & & 0.08*** & & 0.04* & 0.09*** & 0.09*** & 0.05*** & \\

\midrule\multicolumn{8}{@{}l@{}}{\emph{Country (baseline: Germany (DEU))}}\\
DEU - CHN & 0.46*** & 0.38*** & 0.23*** & 0.51*** & 0.71*** & 0.79*** & 0.54*** & 0.68*** \\
DEU - GBR & -0.11* & 0.33*** & & 0.19*** & 0.23*** & 0.35*** & 0.15*** & 0.18*** \\
DEU - IND &  & 0.72*** & & 0.48*** & 0.67*** & 0.84*** & 1.02*** & 0.60*** \\
DEU - ISR &  & & -0.12** & 0.12* & & 0.23*** & 0.11** & \\
DEU - ITA & -0.14** & 0.18*** & -0.24*** & 0.19*** & 0.27*** & 0.33*** & 0.19*** & \\
DEU - MEX & & 0.43*** & -0.17*** & 0.41*** & 0.47*** & 0.64*** & 0.73*** & 0.35*** \\
DEU - POL & 0.14** & 0.24*** &  & 0.31*** & 0.31*** & 0.57*** & 0.47*** & 0.33*** \\
DEU - SAU & 0.17*** & 0.57*** & -0.11* & 0.45*** & 0.53*** & 0.80*** & 0.84*** & 0.67*** \\
DEU - SWE & 0.12** & & & 0.18*** & 0.15*** & 0.34*** & 0.10** & \\
DEU - USA & 0.12** & 0.22*** & 0.18*** & 0.35*** & 0.28*** & 0.37*** & 0.23*** & 0.37*** \\
DEU - ZAF &  & 0.20*** & -0.30*** & 0.51*** & 0.34*** & 0.54*** & 0.52*** & 0.52*** \\

\bottomrule
\end{tabular*}
\end{table*}
\end{center}

%% file: sections/07_discussion.tex
\section{Discussion}
\label{sec:discussion}

We identified several factors that impact users' security and privacy misconceptions and provide insights to the prevalence of security and privacy (mis)conceptions around the world. 
Our systematic analysis of misconceptions can serve as a foundation for research investigating how these misconceptions are formed and how they can be overcome. 
Our results show that usable security and privacy research in different countries is crucial, as we found significant differences between participants in different countries. Specifically, we found that differences were strongest between Western and non-Western countries.

\subsection{Misconceptions}
We found rather moderate (dis)agreement with most misconceptions across all countries. 
Mean values for misconception scores ranged around the neutral middle~($3$) of our 5-point rating scale. 
This shows that users are uncertain about many aspects of their digital security and privacy rather than having misconceptions. 
This could be due to the complexity of the topic, but also due to confusing and unclear information accessible to users. 
Our covariance analyses supported this interpretation, as seeking information on digital security was a positive factor for having misconceptions across topics. 

However, we found clear misconceptions on a number of topics, some of which are present in all countries and others only in some countries. 
For example, we have found the myth that passwords need to be changed regularly to increase security persists around the world, although security specialists no longer recommend this, but rather advise to change passwords only when the account has been compromised~\cite{nist-17-sp800-63b, mayer-17-pw-misconceptions}. This misconception can potentially be harmful for users as regular password changes can result in weak passwords or repeated reuse of (weak, i.e., easily guessable) passwords. Thus, user accounts and data are easier to hack. \changes{We believe that this misconception exists, as regular password change was recommended for a long time and habit changes are difficult. Furthermore, this advice is probably still given to users (see\cite{mayer-17-pw-misconceptions}) and might thus be still prevalent around the world.}

Another example for a clear misconception found worldwide is that clicking on a link can be very dangerous and will surely result in an infection with malware. 
This shows the underestimation or the lack of awareness or knowledge about the security most browsers and modern operating systems have already built in by default. 
For example, Windows includes an anti-malware component that is active by default and updated automatically, hence only clicking on a link is usually not sufficient to install malware on a computer.
Here, the download and further interaction with a file would be dangerous.
However, this misconception probably only leads to overcautious behavior, and thus does not put users in more danger. \changes{To never click on a link is advice experts recommend for users~\cite{reeder-17-152-simple-steps} -- probably due to the before mentioned reasons -- which might be why this believe is present around the world.}

Especially participants from India, but also from other countries (\eg, Mexico, Saudi Arabia and China, to a lesser degree), believe that HTTPS is an indicator for the trustworthiness of a website. 
This misconception bears the risk of users entering login or bank credentials on a malicious website. 
Thus, users should \changes{be informed} that HTTPS only indicates a secure connection and is not an indicator of the trustworthiness or authenticity of a website.

We have also found that participants, especially in Saudi Arabia but also in other countries (with the exception of the US and Sweden), confuse encryption and authentication, thus mistaking a second authentication factor for an encryption layer. 
Here, our results confirm similar qualitative findings by Krombholz~\etal~\cite{krombholz-19-mental-models-https} and even conclude that this misconception is geographically widespread. 
This demonstrates the worldwide demand for sound advice sources and education materials on this topic. 
As two-factor-authentication becomes more prevalent, services that use this authentication method should educate their users to avoid misconceptions and potential resulting risks.

For some misconceptions topics such as passwords, device security or malware, our model explains more variance than for others, like VPN and Wi-Fi. 
This could be due in part to irrelevant or missing predictors for these topics, such as for HTTPS~(Q10) or Wi-Fi~(Q11), where only eight respectively nine of the 14 predictors were significant. 
Future work could identify which factors better predict (mis)conceptions or knowledge on these topics. Therefore, advice and education could not only target misconceptions, but also target these factors.  

\subsection{Factors Influencing Misconceptions}
The country of residence was the most effective predictor for misconceptions across all topics, except for Wi-Fi. We found especially large differences between German participants and participants from non-Western countries across all misconceptions. The differences of country in its predictive power were somewhat smaller for German participants and those from other Western countries, especially for those from Israel.\footnote{Reflecting on Huntington~\cite{huntington-96-civilizations}, we consider Israel as very close to the Western world and thus a part of it.} This is in line with prior findings, pointing out that the research community should not view results from Western countries as the ``norm'' and results from non-Western countries as ``exotic''~\cite{kou-18-titling}. Similarly to other cross-cultural studies on privacy and security (\eg, \cite{wang-11-concerns, sawaya-17-behavior, harbach-16-keep-on}), our findings on misconceptions show that results differ across cultures \changes{and} that results from Western countries are rather alike but differ in many cases from those of non-Western countries.

Counterintuitively, we found that participants with technical backgrounds (student, degree, or job in the field of computing), were significantly more likely to believe in misconceptions about HTTPS, VPN, passwords, device security and malware. However, a related study found similar results. When investigating computer science master students' security knowledge, Chaudhary~\etal found that these future IT professionals hold dangerous misconceptions~\cite{chaudhary-15-online-sec}. When studying knowledge of the Internet and security behavior, Kang~\etal~\cite{kang-15-data-everywhere} also did not find technical education to be a predictor for privacy and security behavior. We can only guess \changes{that one} reason for this might be the great amount of security advice that even IT experts fail to prioritize~(see \cite{reeder-17-152-simple-steps}) and thus misconceptions arise.   
The finding that seeking information on digital security positively influenced believing in misconceptions was also counterintuitive. Again, we assume that this may be due to the confusing amount of security and privacy advice. 

Chaudhary~\etal also suggest that encounters with threats and crimes might help and predict secure behavior. Similarly, Kang \etal suggests security and privacy behavior to be predictable by experiences. We come to the same conclusion, as experiences with cybercrime were a significantly negative, and thus hindering, predictor for believing in misconceptions across many topics (except for E2EE, Wi-Fi, passwords). 

The amount of used devices was a significant predictor for misconceptions related to E2EE, Wi-Fi and VPN, with using more devices positively influenced believing in these misconceptions. The effects of this predictor were comparatively large. This indicates that the more devices people use, the more misunderstandings especially regarding to Wi-Fi security arise. Especially when taking mobile devices in considerations this seem reasonable, as users are warned about using (open) Wi-Fi in locations like coffee shops~\cite{reeder-17-152-simple-steps}. Thus, they might be confused about under which circumstances using Internet on their mobile devices is safe and what steps they have \changes{to undertake} to make it safe. 

Kang~\etal~\cite{kang-15-data-everywhere} found a positive correlation between the number of threats participants named and measures they took to stay save and conclude awareness to be a predictor for security measures. We find that the amount ($>0$) of security measures taken is a hindering factor for believing in misconception across topics. The result that a greater protection importance for data and accounts was associated with less holding misconceptions points in the same direction. People who want to protect their data and already implement proactive measures are less likely to hold misconceptions. 
Kang~\etal also found that attitudes like ``I have nothing to hide'' discourage people from taking security measures. We found that attitudes in this direction positively influence believing in misconceptions, across topics, with highest effects for attitudes related to statements such as ``nobody is interested in my data.''

\subsection{Practical Implications for Future Work}
Our results reveal cross-cultural differences in security and privacy misconceptions, with more differences between non-Western and Western countries. Future work should thus investigate these differences and reasons for these differences, as well as conduct studies not primarily in Western countries. Based on these differences, we see the opportunity to study \textit{what contributes to effective communication of accurate understanding of security and privacy}, as well as \textit{what contributes to misconceptions} by studying user learning behavior, public outreach and education in those countries where misconceptions were especially low or high. 

This future work could make comparisons both across misconceptions, as well as across countries or clusters of similarly-behaving countries.
Researching how to debunk specific misconceptions could also be a (future) perspective. This could help users to convert misconceptions into understanding. For example, for debunking encryption misconceptions, Schaewitz \etal~\cite{schaewitz-21-e2ee} recommended trust building measures, like telling users something is encrypted.

%% file: sections/08_conclusion.tex
\section{Conclusion}
\label{sec:conclusion}
We reported on a large-scale quantitative online survey of security and privacy (mis)conceptions around the world. We surveyed \changes{$n = 12,351$} participants in $12$ countries on \changes{four} continents. A key contribution of this paper is an overview of factors that influence misconceptions across security and privacy topics. Regarding these factors, we found country of residence to be the strongest predictor for holding misconceptions. We identified the greatest differences between non-Western and Western countries, demonstrating the need for region-specific research on usable security an privacy. 
However, while we did find some specific misconceptions to be present across different countries, like the importance of regular password changes for security reasons, we generally did not observe many outright misconceptions. For the large part of misconceptions across topics, we mainly identified uncertainty. 

Our work lays the foundation for future work investigating misconceptions of participants per country in more depth and research on how to debunk specific misconceptions.
Our results show that it is also important to research (other) factors that might influence (mis)conceptions, like technology readiness or Internet literacy. Thus, advice and educational material could target influencing factors and misconceptions.

\begin{acks}
We want to thank all participants of our study. We would like to thank Carina Wiesen, Jennifer Friedauer, Annalina Buckmann, Maximilian Golla, and so many more for their help with this paper. 
The research was primarily funded by the Deutsche Forschungsgemeinschaft (DFG, German Research Foundation) under Germany’s Excellence Strategy -- EXC 2092 CASA -- 390781972 and also (partly) by the PhD School ``SecHuman -- Security for Humans in Cyberspace'' by the federal state of NRW, Germany.
\end{acks}

%% file: appendix/CountryStudies.tex
\section{Security \& Privacy Studies in Different Countries}
\label{appendix:StudiesByCountries}

\newcolumntype{M}[1]{>{\centering\arraybackslash}m{#1}}

\begin{table}[htbp]
\small
 \Description[SecurityAndPrivacyStudiesInDifferentCountries1]{This table has has 3 rows. It shows studies from different areas of security \& privacy related with our questionnaire and the countries in which the participants were recruited. This table continues on the next side}
    \caption{Relevant studies for our questionnaire and the countries in which they recruited participants. (1/2) \changesAdded{Added this table.}}
    \label{appendix:tab:otherStudies}
\begin{tabularx}{\textwidth}{M{2cm}M{8cm}M{4cm}}
\toprule
\textbf{Topic} & \textbf{Related Research} & \textbf{Countries} \\
\midrule
Cross-Cultural Studies on Privacy and Security &     
\begin{enumerate}
    \item Who Is Concerned about What? A Study of American, Chinese and Indian Users’ Privacy Concerns on Social Network Sites~\cite{wang-11-concerns}
    \item Self-Confidence Trumps Knowledge: A Cross-Cultural Study of Security Behavior~\cite{sawaya-17-behavior}
    \item Keep on Lockin' in the Free World: A Multi-National Comparison of Smartphone Locking~\cite{harbach-16-keep-on}
    \item A Cross-Cultural and Gender-Based Perspective for Online Security: Exploring Knowledge, Skills and Attitudes of Higher Education Students~\cite{chaudhary-15-online-sec}
\end{enumerate}
&
\begin{enumerate} 
    \item US, China, India
    \item China, France, Japan, Russia, South Korea, the US, and the United Arab Emirates
    \item Australia, Canada, Germany, Italy, Japan, Netherlands, the UK, and the US
    \item China, Finland, Pakistan, Nepal, Iran, England, Vietnam
\end{enumerate} 
\\
\midrule
Cybercrimes \& Threats &
\begin{enumerate}
    \item Digitalbarometer 2020: Bürgerbefragung zur Cyber-Sicherheit [German]~\cite{zindler-20-digitalbarometer}
    \item ENISA Threat Landscape 15 Top Threats in 2020~\cite{enisa-20-threats}
    \item Africa Cyber Security Report 2016~\cite{serianu-16-threats}
\end{enumerate}
& 
\begin{enumerate}
    \item Germany
    \item European Union 
    \item Africa
\end{enumerate}
\\
\midrule
Digital Security Measures &
\begin{enumerate}
    \item 152 Simple Steps to Stay Safe Online: Security Advice for Non-Tech-Savvy Users~\cite{reeder-17-152-simple-steps}
\end{enumerate}
&
\begin{enumerate}
    \item Almost half from US, but also from UK, Germany, Australia, Japan, India, Israel, and South-Africa
\end{enumerate}
\\
\midrule
Studies focusing on general awareness of users  &
\begin{enumerate}
    \item My Data Just Goes Everywhere: User Mental Models of the Internet and Implications for Privacy and Security~\cite{kang-15-data-everywhere}
    \item Awareness, Adoption, and Misconceptions of Web Privacy Tools~\cite{story-21-web-privacy}
    \item End User and Expert Perceptions of Threats and Potential Countermeasures~\cite{anell-20-threats}
\end{enumerate}
&
\begin{enumerate}
    \item US 
    \item US 
    \item Germany, Italy, Switzerland, and Portugal (all spoke German) 
\end{enumerate}
\\
\midrule
E2EE communication (Q9) &
\begin{enumerate}
    \item Obstacles to the adoption of secure communication~\cite{abu-salma-17-communication}
\end{enumerate}
&
\begin{enumerate}
    \item UK
\end{enumerate}
\\
\midrule
HTTPS (Q10) &
\begin{enumerate}
    \item If HTTPS Were Secure, I Wouldn’t Need 2FA”- End User and Administrator Mental Models of HTTPS~\cite{krombholz-19-mental-models-https}
    \item Awareness, Adoption, and Misconceptions of Web Privacy Tools~\cite{story-21-web-privacy}
\end{enumerate}
&
\begin{enumerate}
    \item Austria \& Germany,
    \item US
\end{enumerate}
\\

\bottomrule
\end{tabularx}
\end{table}

\begin{table}[htbp]
 \Description[SecurityAndPrivacyStudiesInDifferentCountries2]{This table has has 3 rows. It shows studies from different areas of security \& privacy related with our questionnaire and the countries in which the participants were recruited. This table starts on the previous page.}
    \caption{Relevant studies for our questionnaire and the countries in which they recruited participants. (2/2) \changesAdded{Added this table.}}
    \label{appendix:tab:otherStudies2}
\begin{tabularx}{\textwidth}{M{2cm}M{8cm}M{4cm}}
\toprule

\textbf{Topic} & \textbf{Related Research} & \textbf{Countries} \\

\midrule
WiFi (Q11) &
\begin{enumerate}
    \item When I am on Wi-Fi, I am fearless: privacy concerns \& practices in everyday Wi-Fi us~\cite{klasnja-09-wifi}
\end{enumerate}
&
\begin{enumerate}
    \item US
\end{enumerate}
\\

VPN (Q12) &
\begin{enumerate}
    \item Awareness, Adoption, and Misconceptions of Web Privacy Tools~\cite{story-21-web-privacy}
\end{enumerate}
& 
\begin{enumerate}
    \item US
\end{enumerate}
\\
\midrule
Password and Login Processes (Q13) &
\begin{enumerate}
    \item Addressing Misconceptions About Password Security Effectively~\cite{mayer-17-pw-misconceptions}
\end{enumerate}
&
\begin{enumerate}
    \item No participants, meta-analysis
\end{enumerate}
\\
\midrule
Device Security (Q14) &
\begin{enumerate}
    \item End User and Expert Perceptions of Threats and Potential Countermeasures~\cite{anell-20-threats}
    \item Addressing Misconceptions About Password Security Effectively~\cite{mayer-17-pw-misconceptions}
    \item I cannot do anything: User’s Behavior and Protection Strategy upon Losing, or Identifying Unauthorized Access to Online Account~\cite{al2020cannot}
    \item 152 simple steps to stay save online: Security Advice for non-tech-savy Users~\cite{reeder-17-152-simple-steps}
\end{enumerate}
&
\begin{enumerate}
    \item Germany, Italy, Switzerland, and Portugal (all spoke German)
    \item No participants, meta-analysis
    \item Bangladesh, Turkey, and US
    \item Almost half from US, but also from UK, Germany, Australia, Japan, India, Israel, and South Africa
\end{enumerate}

\\
\midrule
Malware and Deception on the Internet (Q15) &
\begin{enumerate}
    \item \enquote{Malware Myth} in Cyberdanger~\cite{willems-19-myths}
    \item Security education against phishing: A modest proposal for a major re-think~\cite{kirlappos-12-phishing}
    \item Modelling User-Phishing Interaction~\cite{dong-08-phishing}
\end{enumerate}
&
\begin{enumerate}
    \item The Netherlands, Belgium, Germany, Switzerland, Austria, the United Kingdom, Russia, Spain, Italy, Poland, and the US
    \item UK 
    \item no participants
\end{enumerate}
\\
\midrule
Private Browsing (Q16) &
\begin{enumerate}
    \item Your Secrets Are Safe: How Browsers’ Explanations Impact Misconceptions About Private Browsing Mode~\cite{wu-18-private-browsing}
\end{enumerate}
&
\begin{enumerate}
    \item US
\end{enumerate}

\\

\bottomrule

\end{tabularx}
\end{table}

%% file: appendix/Model_Variables.tex
\section{List of Excluded Variables per Model}
\label{appendix:modelVariables}
\begin{itemize}
    \item E2EE~(Q9): no variables excluded
    \item HTTPS~(Q10): gender, education, Q17, Q19 -- complicated 
    \item Wi-Fi~(Q11): Q7, Q8, Q17, Q19 -- E2EE, Q19 -- interest, Q25
    \item VPN~(Q12): Q17, Q19 -- complicated, Q19 -- E2EE, Q22
    \item Passwords~(Q13): no variables excluded
    \item Device Security~(Q14): no variables excluded
    \item Malware~(Q15): no variables excluded
    \item Private Browsing~(Q16): Q19 -- complicated
\end{itemize}

%% file: appendix/tab-reliability-predictors.tex
\section{Reliability Coefficient Cronbach's Alpha}
\label{appendix:reliabilitytab}
\begin{table}[H]
    \centering
    \small
    \Description[CovarianceAnalysis]{This table has 7 rows. Every row shows a predictor scale of our covariance analysis and the corresponding reliability coefficient Cronbach's alpha ($\alpha$).}
    \caption{Reliability coefficient Cronbach's alpha ($\alpha$) for all predictor scales used. \changes{Added this table.}}
    \label{appendix:predictorreliability}
    \vspace{-0.3cm}
\begin{tabular}{@{}lc@{}}
    \toprule
    Misconception Topic & $\alpha$ \\
    \midrule
    Q17. Prevention & 0.89 \\
    Q18. Concerns & 0.96 \\
    Q19. Attitudes E2EE  & 0.51 \\
    Q19. Attitudes Interest & 0.86 \\
    Q19. Attitudes Complicated  & 0.80 \\
    Q21. Data Protection & 0.93 \\
    Q22. Potential Attackers & 0.87 \\
    \bottomrule
\end{tabular}
\end{table}

%% file: appendix/questionnaire.tex
\definecolor{headlines}{HTML}{58429B}
\definecolor{info}{HTML}{00AEB3}
\definecolor{description}{HTML}{103778}

\section{Questionnaire -- Survey on Citizens' Digital Security}
\label{appendix:onlinesurvey}
\renewcommand{\labelitemi}{$\bullet$}
\small
\noindent\textbf{{\color{headlines}Welcome Text}}
~\newline
Increasing digitalization in all areas of life leads to more and more people being online and shifting processes from the offline to the online world (\eg, with online banking).
What are the experiences of internet users? How do they perceive different risks and security measures? We would like to answer these questions with this study.
Based on these insights, we aim to develop need-based offers and materials to increase the digital security of the population. You can provide a valuable contribution with your participation.\\

\noindent\textbf{{\color{headlines}Consent -- Data Privacy Statement}}
~\newline
Thank you for your interest in our study.\\
\textbf{Purpose:} The purpose of the study is to gather internet users' experiences of digital security, and how they evaluate various online risks and security measures. Our results will provide a basis for developing communication and training materials that answer people's questions about digital security, and enable them to manage it. You can provide a valuable contribution with your participation.\\
\textbf{Duration:} Participation in the study is expected to take 20~minutes. You are not subject to any anticipated risks by participating.
Please answer the survey as honestly as possible. You may stop at any time if you no longer wish to participate in the study, as long as you have not submitted your responses or these have not been evaluated.\\
\textbf{Data Protection:} Your responses to this study are stored in anonymized form in a way which will not reveal your identity. No data will be passed on to third parties. By starting this questionnaire you consent to data collection for the purposes of conducting this study. Your personal data is processed based on Article 6(1)a of the GDPR. 
You have the right to revoke your consent to the data processing at any time as well as to request information, correction, processing restrictions and deletion of the data stored about you. To exercise these rights, please contact the email address listed below. The responsible supervisory authority is \textit{{\color{info}[blinded]}}.
If you have additional questions about data protection, please contact \textit{{\color{info}[blinded for anonymous review]}}.

\begin{enumerate}
    \item[\textbf{Q0}:] \textbf{Consent}. \textit{{\color{info}[checkbox]}}
\begin{itemize}
    \item \emph{I confirm that I accept the participation conditions for this study.}
\end{itemize}
\end{enumerate}

\noindent\textbf{{\color{headlines}Internet Usage}}
~\newline
First, we would like to ask you some questions about your internet usage.

\begin{enumerate}
    \item[\textbf{Q1}:] \textbf{Which of the following devices to you use in your daily life?} \textit{{\color{info}[multiple choice]}}
\begin{itemize}
    \item \emph{Smartphones; Static PCs / Desktop PCs; Laptops; Tablets; Voice Assistants or Smart Speakers (\eg, Alexa, Amazon Echo); Wearables (\eg, fitness trackers, smartwatches or other computer technologies that are worn on the body); None of the listed devices \textit{{\color{info}[exclusive]}}}
\end{itemize}
\end{enumerate}

\begin{enumerate}
    \item[\textbf{Q2}:] \textbf{Do you have any smart home devices in your household? If yes, what purpose?} \textit{{\color{info}[matrix question]}}
\begin{itemize}
    \item {\color{description}Description}: The ``smart home'' area includes all networked devices that you use in your living space. For example, systems that automatically open or close windows, doors and shutters -- so-called home automation technology. But smart home also includes household appliances such as refrigerators that keep you informed about their contents or robotic vacuum cleaners. These devices can often be operated from anywhere and many of these devices are connected to the internet.
    \item {\color{description}Items}: \emph{Energy and climate (\eg, ``intelligent'' lights or radiators); Security (\eg, networked alarm systems or video monitoring); Home and garden (\eg, ``intelligent'' shutters, robotic vacuum cleaners)}
    \item {\color{description}Answer Options}: \emph{Yes; No; I am not sure}
\end{itemize}
    \item[\textbf{Q3}:] \textbf{How often do you use the internet for the following purposes?} \textit{{\color{info}[multiple choice]}}
\begin{itemize}
    \item {\color{description}Items}: \emph{Online shopping; Ordering services (\eg, booking travel, ordering food, car sharing); Selling goods or services (\eg, through auctions); Researching information and forming opinions (\eg, reading online newspapers); Uploading and sharing personal content you have created yourself (texts, images, photos, videos, music, software); Expressing opinions (\eg, posts on social media, online comments); Online banking; Communication (email, chat, video conferences etc.); Entertainment (\eg, streaming films, music, online games); Official transactions (\eg, ordering an identity card, tax return); Health services (\eg, electronic patient record, virtual doctor appointment); Map services / navigation; Data storage (cloud services)}
    \item {\color{description}Answer Options}: \emph{Never; Less than once a month; Once a month; Several times a month; Once a week; Several times a week; Every day; Several times a day}
\end{itemize}
    \item[\textbf{Q4}:] \textbf{How often do you use the following communication channels?} \textit{{\color{info}[multiple choice]}}
\begin{itemize}
    \item {\color{description}Items}: \emph{Making telephone calls with a land line; Making telephone calls with a smartphone / mobile telephone; Video calls (\eg, Skype, Zoom, Microsoft Teams); Text messaging (SMS); Messenger services (\eg, WhatsApp, Signal); Social media (\eg, Facebook, Twitter, Instagram); E-mail; Online forums and communities}
    \item {\color{description}Answer Options}: \emph{Never; Less than once a month; Once a month;Several times a month; Once a week; Several times a week; Every day; Several times a day}
\end{itemize}
\end{enumerate}

\noindent\textbf{{\color{headlines}Digital Security}}
~\newline
Now, we would like to ask you some questions on the subject of digital security.

\begin{enumerate}
    \item[\textbf{Q5}:] \textbf{When you think about the subject of digital security, is there anything you are concerned about?} \textit{{\color{info}[free text]}}
\begin{itemize}
    \item {\color{description}Description}: Please state everything that occurs to you. You are welcome to also respond in bullet points.
\end{itemize}
    \item[\textbf{Q6}:] \textbf{How familiar are you with the following terms?}
\begin{itemize}
    \item {\color{description}Description}: For each of the following terms, please state how familiar you are with it.
    \item {\color{description}Items}: \emph{Malware (viruses, worms, spyware, Trojans); Ransomware (extortion software); Phishing; Spear phishing; Two factor authentication (2FA); Biometric authentication process ; Identity theft ; Data leak / data theft; HTTPS; Hard drive encryption; End-to-end encryption; Transport encryption; Browser; Private browser mode (incognito mode); IP address; URL; VPN (virtual private network); Tor network; Ad blocker; (Love) scam / romance scam on the internet; Spam; Cloud}
    \item {\color{description}Answer Options}: \emph{I've never heard of this; I've heard of this but I don't know what it; I know what this is but I don't know how it works; I know how this works; I know very well how this works}
\end{itemize}
    \item[\textbf{Q7}:] \textbf{Have you personally been affected by cybercrime?}
\begin{itemize}
    \item {\color{description}Description}: For each of the following items please state if you have been affected.
    \item {\color{description}Items}: \emph{Malware such as viruses or Trojans; Phishing (spying on confidential data); Ransomware or extortion software; Cyberbullying; Fraud with online shopping; External access to an online account; Cyberstalking; Data abuse (passing on or sale of personal data such as telephone number, address, bank details); Love scam / romance scam on the internet}
    \item {\color{description}Answer Options}: \emph{Yes; No; I prefer not to answer this question}
\end{itemize}
\item[\textbf{Q8}:] \textbf{Where do you look for information on the topic of digital security?}
\begin{itemize}
    \item {\color{description}Description}: From the following information sources, please select all the ones that you use to inform yourself about digital security.
    \item {\color{description}Items}: \emph{Print media; Online news; Social media; Radio / podcasts; Television; Friends  and family; IT security experts; Consumer center, authorities; Other \textit{{\color{info}[free text]}}}
\end{itemize}
\end{enumerate}

\noindent\textbf{{\color{headlines}Misconceptions Digital Security}}
~\newline
Next, you will see a number of statements on the topic of digital security. Please carefully read each statement and state how much you agree with the respective statement.\\
~\newline
\textbf{E2E Messenger}
\begin{enumerate}
    \item[\textbf{Q9}:] \textbf{The following statements refer to communication with messenger that use end-to-end encryption (\eg, WhatsApp, Signal).} \textit{{\color{info}[matrix question]}}
\begin{itemize}
    \item {\color{description}Description}: Please read each statement carefully, and indicate how much you agree with each statement.
    \item {\color{description}Items}: \emph{If my chat messages are protected by end-to-end encryption, then my messages can only be read on my device and by the recipient; nobody else can access and read them in transit; Not even the communication service provider that I use can read my messages if they are protected by end-to-end encryption; If someone has access to my smartphone, then this person can read my messages in the messenger app, despite end-to-end encryption; Because the developers of the messenger service know how the encryption works, they can also read my messages despite end-to-end encryption; The end-to-end encryption in messenger services is not secure because any encryption can be broken; If messages are end-to-end encrypted, they are sent directly from my device to the recipient's device, without any intermediate steps; If messages are end-to-end encrypted, they can also be read by third parties during transmission; If I send messages with end-to-end encryption, nobody knows when and with whom I am communicating; Messages that are sent over the internet are easier to read than text messages that are sent via the telephone network.}
    \item {\color{description}Answer Options}: \emph{1 -- fully disagree; 2 -- mainly disagree; 3 -- neutral; 4 -- mainly agree; 5 -- fully agree; I don't understand the statement}
\end{itemize}
\end{enumerate}
\textbf{HTTPS}
\begin{enumerate}
\item[\textbf{Q10}:] \textbf{Next you will see some  statements about digital security when surfing on the internet. Generally an internet browser (\eg, Firefox, Chrome, Edge, Internet Explorer) is used for this.} \textit{{\color{info}[matrix question]}}
\begin{itemize}
    \item {\color{description}Description}: Please carefully read each statement and indicate how much you agree with the respective statement. If we mention ``HTTPS'' for websites, we mean websites that show a lock symbol in the address bar of your internet browser, like in this illustration: \textit{{\color{info}[image]}}
    \item {\color{description}Items}: \emph{I can identify a fraudulent website (\eg, a fake online shop that wants to capture my data), because no lock symbol is shown in the address bar of the internet browser; If HTTPS is used on a website, my internet provider does not know what I am clicking on the website; HTTPS prevents the website operator from seeing what I am clicking on and viewing on the website; Websites that use HTTPS are trustworthy; If I visit websites that use HTTPS then other people that use my computer cannot see where I have been on the internet.}
    \item {\color{description}Answer Options}: \emph{1 -- fully disagree; 2 -- mainly disagree; 3 -- neutral; 4 -- mainly agree; 5 -- fully agree; I don't understand the statement}
\end{itemize}
\end{enumerate}
\textbf{Wi-Fi}
\begin{enumerate}
    \item[\textbf{Q11}:] \textbf{Next you will see some statements about digital security when surfing on Wi-Fi networks.} \textit{{\color{info}[matrix question]}}
\begin{itemize}
    \item {\color{description}Description}: Please carefully read each statement and state how much you agree with the respective statement.
    \item {\color{description}Items}: \emph{When I use a public Wi-Fi , other devices that are also using this Wi-Fi  (\eg, laptops of other visitors in an internet caf\'{e}) can see what websites I am visiting; When I use a public Wi-Fi, other devices that are also using this Wi-Fi can generally see what data (\eg, passwords, credit card information) that I enter on websites; When I am connected to a public Wi-Fi, it is easy to infect my device with malware; On a public Wi-Fi, attackers can redirect me to specifically prepared websites and record the data that I enter there; When I use a public Wi-Fi, other devices that are also using this Wi-Fi can also read and change my emails.}
    \item {\color{description}Answer Options}: \emph{1 -- fully disagree; 2 -- mainly disagree; 3 -- neutral; 4 -- mainly agree; 5 -- fully agree; I don't understand the statement}
\end{itemize}
\end{enumerate}
\textbf{VPN}
\begin{enumerate}
    \item[\textbf{Q12}:] \textbf{Next you will see some messages about digital security when surfing on the internet with a VPN (virtual private network).} \textit{{\color{info}[matrix question]}}
\begin{itemize}
    \item {\color{description}Description}: Please carefully read each statement and state how much you agree with the respective statement.
    \item {\color{description}Items}: \emph{When I use a VPN, my internet provider can no longer see what websites I visit; A VPN prevents malware from reaching my device; A VPN protects me from entering my passwords or credit card information on dangerous websites; A VPN protects me from unauthorized persons getting access to my device; A VPN is like end-to-end encryption between the website and my device; When I use a VPN, the VPN provider can see what websites I visit; When I use a VPN, the VPN provider can see in principle what data I enter on a website (\eg, passwords, credit card information); Surfing via the Tor network prevents my internet provider from seeing what websites I visit.}
    \item {\color{description}Answer Options}: \emph{1 -- fully disagree; 2 -- mainly disagree; 3 -- neutral; 4 -- mainly agree; 5 -- fully agree; I don't understand the statement}
\end{itemize}
\end{enumerate}
\textbf{Login and Passwords}
\begin{enumerate}
    \item[\textbf{Q13}:] \textbf{Next you will see some statements on the topic of passwords and login processes.} \textit{{\color{info}[matrix question]}}
\begin{itemize}
    \item {\color{description}Description}: Please carefully read each statement and state how much you agree with the respective statement.
    \item {\color{description}Items}: \emph{The security of a password is higher if it includes numbers or special characters as well as letters; To increase the security of a password, it is sufficient to replace letters by numbers, for example to replace an ``i'' with a ``1''; To increase the security of a password, it is sufficient to use a word from a different language; A date of birth is a secure password as long as it isn't my own date of birth; The security of a password only depends on the length of the password; It is important for the security of my user accounts to regularly change the password; Attackers try to guess my password and enter a lot of different passwords manually; Using one strong password to login into different user accounts is perfectly safe; Password managers generate secure passwords that cannot be guessed, even with technical assistance; It is more secure to choose a weaker password that is easy to remember, than to write a strong password down (\eg, a note); This is a control question. Please click on the answer ``4 -- mainly agree'' \textit{{\color{info}[validity check]}}; A password manager that I can use to manage and store all my accounts and passwords is not secure; I have to log in to online banking with two processes so that the connection is encrypted, for example, with a password and TAN (transaction number); If, in addition to entering my password, I have to confirm that I want to login into my email mailbox by mobile phone, it is harder for attackers to get into my email mailbox; Facial recognition to log into my user account is very easy to trick, for example with a photo; If I use, my fingerprint to log in to an Apple or Android smartphone, this is stored with the provider and can be stolen from there; It is easier to steal my fingerprint and use it for authentication on my device than it is to guess my password; Login processes such as fingerprints or facial recognition are imprecise and therefore less secure than passwords.}
    \item {\color{description}Answer Options}: \emph{1 -- fully disagree; 2 -- mainly disagree; 3 -- neutral; 4 -- mainly agree; 5 -- fully agree; I don't understand the statement}
\end{itemize}
\end{enumerate}
\textbf{Security of End Devices}
\begin{enumerate}
    \item[\textbf{Q14}:] \textbf{Next you will see some statements on the topic of digital security of end devices.} \textit{{\color{info}[matrix question]}}
\begin{itemize}
    \item {\color{description}Description}: Please carefully read each statement and state how much you agree with the respective statement.
    \item {\color{description}Items}: \emph{When I enter my laptop password in public, somebody could look over my shoulder and read the password; To protect the data on my laptop even if it is stolen, a hard drive encryption must be used; Even if my laptop is stolen, my data is secure because my user account is protected by a password; Anti-virus software doesn't only protect my PC from viruses, but also protects my online user accounts from attacks; Regular updates are sufficient to protect my device and my data from attacks; I don't need to lock devices,such as my laptop, PC, smartphone etc. -- when I am not using them, because the screen is dark anyway and nobody can read it; It is safer to send sensitive data via a computer than via a smartphone; The PIN for the SIM card is sufficient to protect the data on my smartphone; Strangers cannot access my smart home devices as long as I use a secure password for them.}
    \item {\color{description}Answer Options}: \emph{1 -- fully disagree; 2 -- mainly disagree; 3 -- neutral; 4 -- mainly agree; 5 -- fully agree; I don't understand the statement}
\end{itemize}
\end{enumerate}
\textbf{Malware}
\begin{enumerate}
    \item[\textbf{Q15}:] \textbf{Next you will see some statements on the topic of malware and deception on the internet.} \textit{{\color{info}[matrix question]}}
\begin{itemize}
    \item {\color{description}Description}: Please carefully read each statement and state how much you agree with the respective statement.
    \item {\color{description}Items}: \emph{If I don't discover anything suspect on my computer, then it is not infected with malware; As long as I don't download anything, my PC cannot be infected with malware (even if I visit a risky website); Is it more likely to pick up malware from visiting a porn website than visiting a website on the topic of sport; As long as I don't open a file infected with malware, it can't do any damage; Malware is mostly distributed via USB sticks; If Windows is not installed on my PC, it is more secure from attacks, because attackers do not bother to attack operating systems few people use; Malware can be installed on my device (Laptop/PC) without me noticing it directly; Malware can cause  me no longer being able to view my data, having to pay the attackers money to release it; It is sufficient to look at the sender to check the security of emails before opening; My PC can get infected with malware by clicking on a link; I can click on attached files without concern for an email that is addressed to be directly; As long as a website looks official, I can enter my login data without concern; The email could be risky if the sender name and email address are not the same; The text on a link shows me what site you will end up on if you click on it; As long as I know the sender of an email then I don't have to worry about the email containing viruses; Links in emails can lead to fake websites to gather my login data.}
    \item {\color{description}Answer Options}: \emph{1 -- fully disagree; 2 -- mainly disagree; 3 -- neutral; 4 -- mainly agree; 5 -- fully agree; I don't understand the statement}
\end{itemize}
\end{enumerate}
\textbf{Surfing in Private Browsing Mode}
\begin{enumerate}
    \item[\textbf{Q16}:] \textbf{Next you will see some statements about digital security when surfing in private browsing mode (also called incognito mode).} \textit{{\color{info}[matrix question]}}
\begin{itemize}
    \item {\color{description}Description}: Please carefully read each statement and state how much you agree with the respective statement.
    \item {\color{description}Items}: \emph{The private browser mode encrypts my data; The private browser mode prevents my internet provider from seeing what websites I visit; The private browser mode protects me from other people using my device from being able to track my activities; The private browser mode prevents malware from reaching my device; The private browser mode has the same protective effect as an ad blocker, that is, advertising is blocked on a website; The private browser mode does not prevent website operators from being able to see my IP address.}
    \item {\color{description}Answer Options}: \emph{1 -- fully disagree; 2 -- mainly disagree; 3 -- neutral; 4 -- mainly agree; 5 -- fully agree; I don't understand the statement}
\end{itemize}
\end{enumerate}

\noindent\textbf{{\color{headlines}Digital Security Concerns}}
~\newline
Next, you will see a number of statements relating to digital security. Please carefully read each statement and state how much you agree with the respective statement.

\begin{enumerate}
    \item[\textbf{Q17}:] \textbf{How important is it to you to \textbf{prevent}\ldots} \textit{{\color{info}[matrix question]}}
\begin{itemize}
    \item {\color{description}Items}: \emph{malware such as viruses or Trojans from reaching your devices (PC, laptop, smartphone)?; your data (such as login data) from being spied on?; you from no longer being able to view your data and having to pay blackmailers money to view your data?; you from being insulted online? (cyberbullying); you from being a victim of fraud, for example, when shopping online?; unauthorized persons from having access to your online accounts?; unauthorized persons from gaining access to your personal data? your digital messages, such as emails, being accessed and read by third parties?; you becoming a victim of cyberstalking?; your passwords from being guessed by unauthorized persons?; your devices (PC, laptop, smartphone) from being spied on?; you from entering your login data on fraudulent websites?; friends or family with access to your devices (PC, laptop, smartphone) being able to see your browser history?; advertisers from being able to see what websites you visit?; the contents of your messages from being read by communication service providers, \eg, the messenger service?}
    \item {\color{description}Answer Options}: \emph{1 -- not important; 2 -- a little important; 3 -- moderately important; 4 -- quite a bit important; 5 -- very important ; I don't understand the question}
\end{itemize}
    \item[\textbf{Q18}:] \textbf{For each of the following statements, please state how concerned you are. How \textbf{concerned} are you\ldots} \textit{{\color{info}[matrix question]}}
\begin{itemize}
    \item {\color{description}Items}: \emph{that when using messenger services your messages could also be read by unauthorized persons?; that the messenger service provider has access to your message contents, such as sent texts or images?; that other people could read your messages despite end-to-end encryption?; that a website could use an illegal mechanism to collect personal information about you?; that when using a public Wi-Fi other devices could see what data (\eg, passwords, credit card information) you enter on websites?; that somebody could track you based on your location?; that somebody could steal your passwords?; that your biometric data could be abused, \eg, your fingerprint to unlock the mobile phone?; that one of your passwords is easy to crack or guess?; that sensitive data on your computer is not secure enough? (\eg, through backups or firewalls); that someone could get the password for your computer by watching you enter it?; that, if your computer is stolen, unauthorized persons could have access to your sensitive data and passwords?; that your computer could be affected by malware and you would no longer be able to open your files because of it?; that your computer could be affected by malware and is therefore no longer usable?; that your computer could be affected by malware and therefore unauthorized persons have access to your data?; that your computer could have a virus that you don't know about?; that unauthorized third parties could have access to your data?; that networked devices such as voice assistants (\eg, Alexa, Siri) inadvertently gather, store and forward personal data?; that voice assistants, such as Alexa or Siri, inadvertently listen to everything you say?}
    \item {\color{description}Answer Options}: \emph{1 -- not concerned ; 2 -- a little concerned ; 3 -- moderately concerned; 4 -- quite a bit concerned; 5 -- very concerned; I don't understand the question}
\end{itemize}
    \item[\textbf{Q19}:] \textbf{For each of the following statements, please state how much you agree.} \textit{{\color{info}[matrix question]}}
\begin{itemize}
    \item {\color{description}Items}: \emph{I am not rich or famous, so nobody is interested in accessing my personal data;I do not believe that anyone is interested in reading my messages (\eg, emails, chats); I have nothing to hide, therefore it is not important to me whether my messages are encrypted or not; I consciously use communication services (\eg, messenger services) that use end-to-end encryption, because I don't want unauthorized persons to be able to read my messages; I don't need strong passwords, because my data is not interesting to attackers; People who use the private browser mode have something to hide. Wi-Fi at home is more secure than public Wi-Fi; Encryption is only for people who are paranoid. Encryption has more advantages than disadvantages; Encryption is dangerous, because I can irretrievably lose my data; Encryption is bad because it is used by hackers and criminals (\eg, for illegal activities); Encryption is useful to ensure protection of personal data; Digital security is complicated; Products with a high level of security are often difficult to use; Secure programs or applications are often difficult to use; Programs and services should be secure. It is not my job to take care of security; Regardless of what I do, I am powerless against skilled attackers and hackers. I don't want to have to deal with digital security; Digital security is annoying.}
    \item {\color{description}Answer Options}: \emph{1 -- fully disagree; 2 -- mainly disagree; 3 -- neutral; 4 -- mainly agree; 5 -- fully agree; I don't understand the statement}
\end{itemize}
\end{enumerate}

\noindent\textbf{{\color{headlines}Digital Security Risk and Measures}}
~\newline
You are almost done! Last but not least, we would like to learn which measures you take to stay safe on the internet.

\begin{enumerate}
    \item[\textbf{Q20}:] \textbf{What measures do you use for your digital security?}  \textit{{\color{info}[matrix question]}}
\begin{itemize}
    \item {\color{description}Description}: Please click on all the measures you use for your digital security.
    \item {\color{description}Items}: \emph{(Regular) updates of the operating system and other programs; (Regular) backups on an external hard drive; (Regular) backups to the cloud; Anti-virus software; Firewall; Ad blocker; Anti-tracking tools; Password manager; End-to-end encryption for messages; PIN, password or biometric processes to lock and unlock your devices (laptop, smartphone, tablet); Two factor authentication; Tor network; VPN (virtual private network); None \textit{{\color{info}[exclusive]}}}
\end{itemize}
    \item[\textbf{Q21}:] \textbf{How important is it for you to protect the following data on the internet (\eg, from external access and theft)?} \textit{{\color{info}[matrix question]}}
\begin{itemize}
    \item {\color{description}Items}: \emph{Your full name; Address (home address); Your personal telephone numbers; Your contacts; Your personal photos; Message threads, for example, from chats and emails; Location and movements, \eg, GPS data, your jogging route; The amount of your salary or earnings; ID, such as identity card and driving license; Insurance documents; Delivery notes and invoices; IBAN / BIC and account details; Health data; Biometric data, such as fingerprints; Passwords}
    \item {\color{description}Answer Options}: \emph{1 -- not important; 2 -- a little important; 3 -- moderately important; 4 -- quite a bit important; 5 -- very important ; I don't understand the question}
\end{itemize}
    \item[\textbf{Q22}:] \textbf{How likely is it that the following groups of people pose a risk to your digital security (\eg, unauthorized access to your personal data, stalk you online or restrict your access to digital services)?} \textit{{\color{info}[matrix question]}}
\begin{itemize}
    \item {\color{description}Items}: \emph{Family members; Friends and acquaintances; Work colleagues; Officials from [country] (such as police, secret services and the government); Officials from other countries (such as police, secret services and the government); Private sector companies; Criminals who want to get rich from your data; Hackers who gain unauthorized access to data and devices, for fun.}
    \item {\color{description}Answer Options}: \emph{1 -- not likely; 2 -- a little likely; 3 -- moderately; 4 -- quite a bit likely; 5 -- very likely}
\end{itemize}
\end{enumerate}

\noindent\textbf{{\color{headlines}Demographics}}
~\newline
Finally, we would like to ask you some more questions about you.

\begin{enumerate}
    \item[\textbf{Q23}:] \textbf{What is your gender?} \textit{{\color{info}[single choice]}}
\begin{itemize}
    \item \emph{Male; Female; Non-binary; Describe yourself: \textit{{\color{info}[free text]}}; I prefer not to answer this question}
\end{itemize}
    \item[\textbf{Q24}:] \textbf{What is your highest level of education?} \textit{{\color{info}[single choice]}}
\begin{itemize}
    \item \emph{No school leaving certificate; Secondary school (primary school) or equivalent leaving certificate; High school (O level) or equivalent leaving certificate; A level, vocational high school / general or university entrance qualification; Occupational or vocational training / apprenticeship; Completion of a technical college or administrative or professional academy; Bachelor's degree; Diploma university course or masters (including: teaching position, state examination, Master's course, artistic or comparable courses of study); PhD/doctorate; I prefer not to answer this question}
\end{itemize}
    \item[\textbf{Q25}:] \textbf{Do you have practical experience in the informatics, computer technology or information technology fields (\eg, through your job or education background)?} \textit{{\color{info}[single choice]}}
\begin{itemize}
    \item \emph{Yes; No; I prefer not to answer this question}
\end{itemize}
    \item[]\hspace{-7.7mm}\textbf{Q\_Hidden}: \textbf{Country} \textit{{\color{info}[hidden question]}}
\begin{itemize}
    \item \emph{Chinese; German; Indian; Israeli; Italian; Mexican; Polish; Arabian; South African; Swedish; British; American}
\end{itemize}
    \item[\textbf{Q26}:] \textbf{Do you have an immigration background?} \textit{{\color{info}[single choice]}}
\begin{itemize}
    \item {\color{description}Description}: People with an immigration background are defined as people who were not born as a \textit{{\color{info}[country]}} citizen or who have at least one parent who was not born as a \textit{{\color{info}[country]}} citizen.
    \item {\color{description}Items}: \emph{Yes, I have an immigration background; No, I don't have an immigration background; I prefer not to answer this question}
\end{itemize}
\end{enumerate}

\noindent\textbf{{\color{headlines}Debriefing}}
~\newline
Thank you very much for participating in our survey. The purpose of the study is to discover what experiences internet users have had concerning digital security and how they evaluate various risks and measures. Your participation helps us to gain knowledge of this so that need-based offers and materials can be developed to increase the digital security of the population.